\documentclass[lettersize,journal]{IEEEtran}
\usepackage{amsmath,amsfonts}
\usepackage{alltt}
\usepackage[ruled,vlined,linesnumbered]{algorithm2e}
\usepackage{array}
\usepackage[caption=false,font=normalsize,labelfont=sf,textfont=sf]{subfig}
\usepackage{textcomp}
\usepackage{graphicx}
\usepackage{stfloats}
\usepackage{url}
\usepackage{verbatim}
\usepackage{threeparttable}
\usepackage{booktabs}
\usepackage{bm}
\usepackage{xcolor}
\usepackage{dashbox}
\newtheorem{definition}{Definition}
\newtheorem{theorem}{Theorem}
\def\BibTeX{{\rm B\kern-.05em{\sc i\kern-.025em b}\kern-.08em
    T\kern-.1667em\lower.7ex\hbox{E}\kern-.125emX}}
\usepackage{balance}

\usepackage{cite}
\usepackage[colorlinks,
linkcolor=blue,
anchorcolor=blue,
citecolor=blue]{hyperref}

\usepackage{orcidlink}
\begin{document}
\title{Privacy-Preserving Federated Learning from Partial Decryption Verifiable Threshold Multi-Client Functional Encryption}
\author{Minjie~Wang$^{\orcidlink{0009-0004-4704-6126}}$,
	Jinguang~Han$^{\orcidlink{0000-0002-4993-9452}}$,~\IEEEmembership{Senior Member,~IEEE,}
	Weizhi~Meng$^{\orcidlink{0000-0003-4384-5786}}$,~\IEEEmembership{Senior Member,~IEEE}
\thanks{
	Minjie Wang is with the School of Cyber Science and Engineering, Southeast University, Nanjing 210096, China (e-mail:220235393@seu.edu.cn).
	
	Jinguang Han is with the School of Cyber Science and Engineering, Southeast University, Nanjing 210096, China and also with the Wuxi Campus, Southeast University, Wuxi 214125, China (e-mail:jghan@seu.edu.cn).
	
	Weizhi Meng is with the School of Computing and Communications, Lancaster University, Lancaster
	University, United Kingdom (e-mail: weizhi.meng@ieee.org).}}

\markboth{Journal of \LaTeX\ Class Files,~Vol.~18, No.~9, September~2020}%
{How to Use the IEEEtran \LaTeX \ Templates}

\maketitle

\begin{abstract}
In federated learning, multiple parties can cooperate to train the model without directly exchanging their own private data, but the gradient leakage problem still threatens the privacy security and model integrity. Although the existing scheme uses threshold cryptography to mitigate the inference attack, it can not guarantee the verifiability of the aggregation results, making the system vulnerable to the threat of poisoning attack. We construct a partial decryption verifiable threshold multi client function encryption scheme, and apply it to Federated learning to implement the federated learning verifiable threshold security aggregation protocol (VTSAFL). VTSAFL empowers clients to verify aggregation results, concurrently minimizing both computational and communication overhead. The size of the functional key and partial decryption results of the scheme are constant, which provides efficiency guarantee for large-scale deployment. The experimental results on MNIST dataset show that vtsafl can achieve the same accuracy as the existing scheme, while reducing the total training time by more than 40\%, and reducing the communication overhead by up to 50\%. This efficiency is critical for overcoming the resource constraints inherent in Internet of Things (IoT) devices.
\end{abstract}


\section{Introduction}

\IEEEPARstart{F}{ederated} Learning (FL) is a distributed machine learning framework, which was first proposed by Google in 2016 \cite{GoogleFL} to solve the growing problem of data privacy. Its core design allows multiple participants to collaborate to train the global model under the coordination of the central aggregation server, while ensuring that the original data of all participants remains localized and never shared. In federated learning, the main role of the central server is to aggregate the model parameters of participants, rather than collect their private data. This model effectively reduces the inherent privacy risks associated with traditional centralized machine learning.

However, FL introduces new privacy vulnerabilities. Adversaries may infer private data from these uploaded updates. For instance, attackers could potentially deduce true labels of input data from leaked gradients or even employ leaked models \cite{inferenceattacks2017}, \cite{inferenceattacks2018}, \cite{inferenceattacks2019}, \cite{zhu2019leakagefromgradients}, \cite{geiping2020inverting}, possibly using Generative Adversarial Networks (GANs) \cite{GAN2017deep}, to reconstruct datasets that approximate the clients' private data, resulting in serious privacy breaches. To address these privacy challenges, several secure aggregation schemes have been proposed to facilitate privacy-preserving federated learning (PPFL), employing diverse primitives. As shown by Table \ref{table:comparison}, existing PPFL approaches exhibit significant diversity in architecture, trust assumptions, and cryptographic methods. Many early schemes \cite{BatchCrypt}, \cite{HE1}, \cite{HybridAlpha}, \cite{revisited}, \cite{DeTrustFLPF}, \cite{DMCFE1} equip a single, Honest-but-Curious (HbC) aggregator with cryptographic techniques, e.g., Homomorphic Encryption (HE) or Functional Encryption (FE). While offering partial protection against gradient inference or "mix-and-match" attacks, these often face the risk of intermediate model leakage and suffer from a single point of failure. Because FE can prevent the aggregator from accessing client local models, but the aggregator can still access the intermediate model via the functional key in federated learning training. On the other hand, although HE protects the intermediate model through direct ciphertext aggregation, it is susceptible to replay attacks and suffers from a single point of failure.

Subsequent works introduced multi-server architectures, such as two-server secure multi-party computation (MPC) \cite{ELSASA} or trust execution environment (TEE) \cite{TEE1}, \cite{ShuffleFL}, often retaining the HbC assumption. More recently, the scheme \textsf{TAPFed} \cite{TAPFed} address adversarial aggregators using multiple-aggregator systems and threshold MCFE. Under a multi-aggregator architecture, the intermediate model cannot be obtained by any independent aggregator, thereby mitigating intermediate model leakage. Furthermore, the threshold mechanism in \textsf{TAPFed} mitigate the single point of failure issue, as the training task can be processed by $t$ active aggregators. This setup is also resilient to collusion attacks from up to $t-1$ aggregators. However, under the assumption of adversarial aggregators, certain aggregators may not adhere to established training protocols and could attempt to tamper with aggregation results. For instance, they might maliciously alter these results to launch poisoning attacks or directly return random values to evade substantial computational overhead. In such circumstances, previous schemes cannot guarantee the validity of the decrypted results furnished by these aggregators, and clients are subsequently unable to discern the validity of these results, thereby failing to achieve the intended training outcomes.

To overcome limitations in prior work, we propose \textsf{VTSAFL}, a novel verifiable threshold secure aggregation scheme for federated learning. \textsf{VTSAFL} ensures aggregation result confidentiality via a threshold mechanism and provides result verifiability for all participants. The scheme is efficient in computation and communication, particularly in large-scale client scenarios.

This work's main contributions include:

\begin{enumerate}
	\item Verifiability of the aggregation results: We construct a partial decryption verifiable threshold multi-client functional encryption scheme, and implement a new privacy preserving federated learning framework called \textsf{VTSAFL}. In this system, each aggregator not only calculates the partial decryption result, but also calculates the corresponding correctness proof using the DLEQ protocol \cite{Chaum1992}. Upon receiving these results and proofs, clients run a verification algorithm to check the validity of the computation performed by each aggregator. If an aggregator provides a malicious or incorrect result, the verification will fail, and its result will be discarded by the clients, thus ensuring the integrity of the global model aggregation
	\item High runtime efficiency and low communication overhead: In our framework, the threshold functional encryption scheme employed can reduce the size of both the functional key and the partial decryption results to a constant level, rather than being proportional to the number of clients. This offers a distinct advantage in large-scale privacy-preserving federated learning tasks involving a substantial number of clients. Additionally, we utilize a multi-secret sharing scheme for the distribution of functional keys, which further helps to minimize the communication overhead.
	\item Comparison and Implementation: We implemented the \textsf{VTSAFL} framework and compared it with existing schemes. Our experiments are carried out on MNIST and CIFAR10 datasets. The final experimental results show that our \textsf{VTSAFL} framework has improved in training time and communication overhead, and is optimized for scenarios with a large number of participants to make it more suitable for IOT applications. The framework not only achieves the same model performance, but also provides the verifiability of aggregate results.
	\item Security Analysis: We present security model for our MCFE scheme and conduct a formal security proof. Subsequently, we analyze the security of the \textsf{VTSAFL} framework, proving its resilience against various attacks.
\end{enumerate}

\begin{table*}
	\begin{threeparttable}
		\caption{Comparison of Different Approaches in Privacy-preserving Federated Learning}
		\label{table:comparison}
		\begin{center}
			\begin{tabular}{%
					>{\centering\arraybackslash}p{1.6cm}                      
					>{\centering\arraybackslash}p{1.85cm} 
					>{\centering\arraybackslash}p{1.4cm} 
					>{\centering\arraybackslash}p{1.7cm} 
					>{\centering\arraybackslash}p{2.4cm} 
					>{\centering\arraybackslash}p{2cm} 
					>{\centering\arraybackslash}p{1.5cm} 
					>{\centering\arraybackslash}p{2cm} 
				}
				\toprule

				& \multicolumn{2}{c}{Aggregator} &  & \multicolumn{3}{c}{Attack Resistance} &\\

				\cmidrule{2-3}
				\cmidrule{5-7}

				\multicolumn{1}{c}{Scheme} & architecture & assumption & Approaches & gradient inference attacks & "mix-and-match" attack &poisoning attack $^\dag$ & Single point of failure\\
				\midrule
				\cite{BatchCrypt},\cite{HE1} & single & HbC & HE & $\checkmark$ & $\times$ & $\times$ &$\times$\\
				\cite{HybridAlpha} & single & HbC & MIFE & $\times$ & $\times$ & $\times$ &$\times$\\
				\cite{revisited} & single & HbC & 2DMCFE & $\checkmark$ & $\checkmark$ & $\times$ &$\times$\\
				\cite{DeTrustFLPF},\cite{DMCFE1} & single & HbC & DMCFE & $\times$ & $\checkmark$ & $\times$ &$\times$\\
				\cite{ELSASA} & two-server & HbC & MPC & $\checkmark$ & $\checkmark$ & $\times$ &$\times$\\
				\cite{TAPFed} & multiple & Adversarial & TMCFE & $\checkmark$ & $\checkmark$ & $\times$ &$\checkmark$\\
				\textbf{Our work} & multiple & Adversarial & VTMCFE & $\checkmark$ & $\checkmark$ & $\checkmark$&$\checkmark$\\
				\bottomrule
			\end{tabular}
			\begin{tablenotes}
				\footnotesize
				
				\item $^\dag$ The defense against poisoning attacks specifically targets those from adversarial aggregators only
			\end{tablenotes}
		\end{center}
	\end{threeparttable}
\end{table*}

\section{Related Work}
\subsection{Multi-input Function Encryption}
Functional encryption (FE) \cite{boneh2011functional},\cite{garg2013candidate} is an example of public key cryptosystem. It enables all parties to encrypt data so that authorized entities holding specific function keys can calculate specific functions on the ciphertext. This calculation shows the output of the function $f(x)$, without disclosing any additional information about the underlying plaintext $x$. The setting of Fe scheme usually involves a trusted institution (TA). TA is responsible for generating master key $msk$ and corresponding master public key $mpk$. $mpk$ is provided to one or more entities that use it to encrypt their input. For the given function $f(\cdot)$, TA uses the $msk$ derived function key $dk_f$. Any designated decryptor with $dk_f$ can calculate the function $f$ on the ciphertext $enc(x) $to obtain the result $f(x)$. It is critical that this assessment is performed directly on encrypted data to ensure the confidentiality of plaintext $x$ throughout the process.

Although the standard FE can operate on large inputs (such as high-dimensional vectors), it essentially assumes that the entire input comes from one party. Therefore, all components of the input vector must be simultaneously provided and encrypted by one entity. For many practical applications, this model is not enough. These applications usually rely on aggregating information from multiple, different and possibly untrusted parties. To address this limitation, Goldwasser et al. \cite{goldwasser2014multi} introduced multiple input function encryption (MIFE). MIFE extends FE to calculate functions on inputs provided by multiple parties. However, MIFE scheme is vulnerable to "hybrid matching" attack. The reason for this vulnerability is that during decryption calculation, ciphertext from different clients (or different time periods) can be combined. For example, consider a scenario where client 1 provides ciphertext for input $\{x_0, x_1\}$ and client 2 provides ciphertext for $\{y_0, y_1\}$. For all possible combinations of $(i, j) \in \{0, 1\}^2$, the calculator may be able to calculate $f(x_i, y_j)$, resulting in serious information leakage. To mitigate this attack, a multi client function encryption (MCFE) \cite{abdalla2018multi},\cite{chotard2018decentralized} is proposed. MCFE enhances MIFE by introducing labels. In the MCFE scheme, each encrypted message is associated with a specific tag. The decryption process is limited, so that the ciphertext can be combined only when sharing the same token, so as to prevent accidental "mixed-and-match" calculation.

\subsection{Privacy Preserving Federated Learning}
In order to achieve privacy preserving federated learning (PPFL), a variety of security aggregation schemes are introduced, which are characterized by the use of different encryption primitives: (1) Differential privacy (DP) reference \cite{liu2020padl},\cite{wu2020value}: DP is a non encrypted method, and users add noise locally to their data before sending the "randomized" version to the aggregator. This process is designed to prevent anyone (including aggregators) from recovering private data. However, a key disadvantage is that the added noise will reduce the prediction accuracy of the obtained intermediate model; (2) Multi-party computation (MPC) \cite{ELSASA},\cite{guo2020v}: most solutions in this domain use secret sharing. Users usually use a one-time pad to mask messages for the aggregator. Decryption requires the aggregator to collect a sufficient number of these shares. Although this method maintains accuracy, it is different from the DP based method in that it needs to increase the interaction between users and aggregators, so as to increase the communication cost and the risk of disconnection; (3) Trust execution environment (TEEs) \cite{TEE1},\cite{ShuffleFL}: the scheme based on trusted execution environment uses Intel SGX, amd PSP and arm TrustZone to create isolated safe enclaves relying on dedicated hardware. However, these solutions face practical limitations, especially the limited availability of required hardware and the limited memory space for secure computing; (4) Homomorphic encryption (HE) \cite{BatchCrypt,HE1}: it allows direct public calculation of encrypted data. In this example, each user encrypts its local model, and then the aggregator can perform arithmetic operations on this model. However, the based solutions are often plagued by computational constraints, making them unsuitable for large-scale security aggregation of model updates. In addition, they also have potential isolation or replay attack vulnerabilities. Other schemes explore different architectures. For example, \textsf{SVFLC} \cite{svflc} proposes a chain aggregation framework, which combines lightweight mask technology to resist collusion attacks, and uses homomorphic hash functions to achieve verifiable aggregation for the server, but cannot resist the poison attack from malicious aggregators. In order to prevent malicious forgery of aggregation results in the cloud, \textsf{FVFL} \cite{fvfl} introduces Lagrange interpolation polynomial and secret sharing to implement an effective verification mechanism. However, this scheme still relies on a centralized server. If the server cannot follow the rules, the whole training process will be difficult to implement.

In recent years, FE based solutions have attracted more and more attention. Compared with the above technologies, the method based on multi-input function encryption (MIFE) has advantages in computational efficiency and communication efficiency. In addition, unlike DP, MIFE based solutions do not compromise model accuracy or rely on other dedicated hardware. Xu et al. \cite{HybridAlpha} represents the first use of MIFE. Its purpose is to prevent a curious aggregator and colluding users from inferring private information. Nevertheless, their proposed scheme is still vulnerable to "Mix-and-Match" attacks and shows vulnerabilities related to the disclosure of intermediate models. Chang et al. \cite{revisited}  designed 2DMCFE and applied it to federal learning. By adding tags to the model ciphertext of the same training batch, they solved the "Mix-and-Match" attack. In addition, in each round of training, clients share a session key to reduce the risk of intermediate model leakage.

Another way to mitigate intermediate model leakage is to use multiple aggregators instead of relying on a single centralized aggregator. Xu et al. First proposed the definition of threshold function encryption, and introduced a new privacy preserving federated learning framework called TAPFed \cite{TAPFed}, in which security aggregation can be achieved through decentralized multiple aggregators, which are independent of each other and do not need peer-to-peer communications. At the same time, the scheme also allows the existence of malicious entities. In this scheme, shamir secret sharing was employed to distribute the functional key. Each aggregator receives only a secret share and computes a partial aggregation result, thereby preventing any single aggregator from directly accessing the intermediate model. However, clients are unable to verify the validity of the aggregation results. This verification is crucial, as the scheme permits the existence of malicious aggregators, which might launch poisoning attacks to disrupt the training process or even return random values instead of valid aggregated results to evade computational costs. Concurrently, in the threshold multi-client functional encryption \cite{TAPFed}, the length of the functional key and the partial decryption results are both proportional to the number of clients. Consequently, this approach is not well-suited for large-scale training tasks that involve a substantial number of clients.

\section{Preliminaries}

\subsection{Notions}\label{s:2.1}
For clarity and convenience, we summarize the main notations used throughout this paper in Table~\ref{tab:notation_summary} including the key parameters and variables related to the system architecture, federated learning process, and the underlying cryptographic scheme.
\begin{table}[htbp]
	\centering
	\caption{Notations Summary}
	\label{tab:notation_summary}
	
	\renewcommand{\arraystretch}{1.2}
	\begin{tabular}{ll}
		\toprule
		\textbf{Notation} & \textbf{Description} \\
		\midrule
		$\lambda$ & The security parameter \\
		$n$ & Total number of clients \\
		$s$ & Total number of aggregators \\
		$t$ & Threshold number of aggregators \\
		$K$ & Maximum number of training rounds \\
		$p_i$ & The $i$-th client \\
		$a_j$ & The $j$-th aggregator \\
		$\theta_{p_i}^{(k)}$ & Local model of client $p_i$ in round $k$ \\
		$\theta_{G}^{(k)}$ & Aggregated global model in round $k$ \\
		$l^{(k)}$ & A unique public label for round $k$ \\
		$pp$ & Public parameters of the cryptosystem \\
		$msk$ & Master secret key \\
		$ek_i$ & Encryption key for client $p_i$ \\
		$dk_j$ & Decryption key share for aggregator $a_j$ \\
		$ct_{p_i}^{(k)}$ & Ciphertext of the local model in round $k$ \\
		$ct_j'$ & A partial decryption result \\
		$\Pi_j$ & A proof of correctness for decryption \\
		PPT & Probabilistic Polynomial Time \\
		$\langle \cdot, \cdot \rangle$ & The inner product operation \\
		$\xleftarrow{\$}$ & Sampling uniformly at random \\
		\bottomrule
	\end{tabular}
\end{table}

\subsection{Computational Assumptions}\label{s:2.2}

\begin{definition}[DDH Assumption]
	\normalfont
	Let $\mathbb{G}$ be a cyclic group of prime order $p$ with generator $g$. The Decisional Diffie-Hellman (DDH) assumption states that no PPT adversary $\mathcal{A}$ can distinguish between the following two distributions with a non-negligible advantage:
	\begin{gather*}
		D_{real} = \left\{ (g^a, g^b, g^{ab}) \mid a,b \overset{\$}{\leftarrow} \mathbb{Z}_p \right\} \\
		D_{rand} = \left\{ (g^a, g^b, g^r) \mid a,b,r \overset{\$}{\leftarrow} \mathbb{Z}_p \right\}
	\end{gather*}
	
	More formally, let $P_{real}$ denote the probability that $\mathcal{A}$ outputs 1 when given an element from $D_{real}$, and $P_{rand}$ be the probability that $\mathcal{A}$ outputs 1 when given an element from $D_{rand}$:
	\begin{align*}
		P_{real} &= \Pr_{a,b \overset{\$}{\leftarrow} \mathbb{Z}_p}[\mathcal{A}(g, g^a, g^b, g^{ab}) = 1] \\
		P_{rand} &= \Pr_{a,b,r \overset{\$}{\leftarrow} \mathbb{Z}_p}[\mathcal{A}(g, g^a, g^b, g^r) = 1]
	\end{align*}
	
	The DDH assumption holds if the advantage, defined as
	\[
	Adv^{DDH}_{\mathcal{A}} = |P_{real} - P_{rand}| \leq \epsilon(\lambda)
	\]
	is negligible.
\end{definition}

\begin{definition}[Multi-DDH Assumption\cite{chotard2018decentralized}]
	\normalfont
	The Multi-DDH assumption, as introduced in \cite{chotard2018decentralized}, posits that for PPT adversary $\mathcal{A}$ operating within $t$, its advantage in discriminating between the distributions $\mathcal{D}_m$ and $\mathcal{D}'_m$ is bounded. These distributions are defined over a group $\mathbb{G}$ as follows:
	\begin{align*}
		\mathcal{D}_m &= \{(X, (Y_j, Z_j)) \mid X, Y_j \overset{\$}{\leftarrow} \mathbb{G}, Z_j = \text{CDH}(X, Y_j)\} \\
		\mathcal{D}'_m &= \{(X, (Y_j, Z_j)) \mid X, Y_j, Z_j \overset{\$}{\leftarrow} \mathbb{G}\}
	\end{align*}
	This advantage is formally bounded by $\mathsf{Adv}_{\mathcal{G}}^{\textsf{ddh}}(t + 4m \times t_{\mathbb{G}})$, where $j = 1, \ldots, m$, and $t_{\mathbb{G}}$ represents the computational cost required for a single exponentiation within the group $\mathbb{G}$.
	
\end{definition}

\subsection{Homogeneous linear recursions}

Homogeneous linear recursions (HLR) are used in multi-secret sharing \cite{mashahdi2022non}.

\begin{definition}[HLR] \label{thm:HLR_solution} \normalfont
	Let $t$ be a positive integer, and $b_0, \ldots, b_{t-1}, z_1, \ldots, z_t \in \mathbb{R}$. A homogeneous linear recursion (HLR) of degree $t$ is defined by:
	\[
	\quad
	\begin{cases}
		w_0 = b_0, w_1 = b_1, \ldots, w_{t-1} = b_{t-1} \\
		w_{i+t} + z_1 w_{i+t-1} + \cdots + z_t w_i = 0, \quad i \geq 0
	\end{cases}
	\]
	The associated auxiliary equation is:
	\[
	x^t + z_1 x^{t-1} + \cdots + z_{t-1} x + z_t = 0.
	\]
\end{definition}

Suppose $\beta_1, \ldots, \beta_l$ are the distinct roots of the auxiliary equation with multiplicities $k_1, \ldots, k_l$ (where $\sum_{j=1}^{l} k_j = t$). The general solution for $w_i$ is:
\[
w_i = \sum_{j=1}^{l} p_j(i) \beta_j^i,
\]
where $p_j(i)$ is a polynomial in $i$ of degree at most $k_j - 1$.

If the auxiliary equation has a single root $\beta$ with multiplicity $t$, the solution simplifies to $w_i = p(i)\beta^i$, where $p(i)$ is a polynomial in $i$ of degree at most $t - 1$.
\subsection{Secret Sharing}\label{s:2.3}

Building upon the properties of HLR, we can construct a Multi-Secret Sharing scheme. This scheme allows a dealer to distribute $m$ secrets among $s$ participants. It is a $(t, s)$-threshold scheme, where any group of $t$ or more participants can reconstruct the secrets. The construction relies on an HLR with a single root of multiplicity $t$. The scheme consists of three algorithms: \textsf{Setup}, \textsf{ShareDistribution}, and \textsf{Reconstruction}.

\textsf{Setup}. The dealer first establishes the public parameters. Given a threshold $t$ and number of participants $s$, the dealer chooses a large prime $p$ and a random element $\alpha \in \mathbb{Z}_p$. Then, the dealer defines the characteristic polynomial $P(x) = (x - \alpha)^t \pmod{p}$ and expands it to obtain the coefficients $a_1, \dots, a_t$. These coefficients define the HLR that will be used. The public parameters are $(p, \alpha, t, s, \{a_i\}_{i \in [t]})$.

\textsf{ShareDistribution}. To share $m$ secrets $(\sigma_1, \dots, \sigma_m) \in \mathbb{Z}_p^m$ where $m < t$, the dealer first constructs the initial state of the sequence $(w_0, \dots, w_{t-1})$. The secrets are embedded as the first $m$ terms, $w_i = \sigma_{i+1}$ for $i \in \{0, \dots, m-1\}$, and the remaining $t-m$ terms are chosen randomly, $w_i \xleftarrow{\$} \mathbb{Z}_p$ for $i \in \{m, \dots, t-1\}$. Then, using the HLR relation, the dealer computes the subsequent terms. The share for the $j$-th participant is $sh_j = w_{t+j-1}$.
\[
w_{i+t} + a_1 w_{i+t-1} + \cdots + a_t w_i \equiv 0 \pmod{p}
\]

\textsf{Reconstruction}. Any group of $t$ participants, holding a set of shares $\{sh_j\}_{j \in S'}$ where $|S'|=t$, can reconstruct the secrets. Based on Theorem~\ref{thm:HLR_solution}, each sequence term satisfies $w_i = q(i) \cdot \alpha^i$ for a polynomial $q(i)$ of degree at most $t-1$. Each participant $j \in S'$ uses their share $sh_j = w_{t+j-1}$ to compute a point on this polynomial: $q(t+j-1) = sh_j / \alpha^{t+j-1}$. With $t$ such distinct points, the group can use Lagrange interpolation to uniquely determine the polynomial $q(i)$. Finally, they recover the secrets by evaluating $q(i)$ at the first $m$ indices: $\sigma_{i+1} = q(i) \cdot \alpha^i \pmod{p}$ for $i \in \{0, \dots, m-1\}$.

\subsection{Discrete Logarithms Equality Protocol}\label{s:2.4}

The Chaum-Pedersen protocol \cite{Chaum1992} is a proof of knowledge for the equality of two discrete logarithms. It allows a prover ($P$) to convince a verifier ($V$) that $y = g^x$ and $t = h^x$ share the same secret exponent $x \in \mathbb{Z}_q$, without revealing $x$.

The protocol operates in a cyclic group $G_q$ of prime order $q$ with distinct generators $g$ and $h$. The proof, $\Pi: \text{Pok} \left\{ (x) : y = g^x \land t = h^x \right\}$, proceeds as follows:

\begin{enumerate}
	\item Commitment: $P$ selects a random $r \in \mathbb{Z}_q$, computes $a_1 = g^r$ and $a_2 = h^r$, and sends $(a_1, a_2)$ to $V$.
	
	\item Challenge: $V$ sends a random challenge $c \in \mathbb{Z}_q$ to $P$.
	
	\item Response: $P$ computes $z = r + cx \pmod{q}$ and sends $z$ to $V$.
	
	\item Verification: $V$ accepts if $g^z = a_1 y^c$ and $h^z = a_2 t^c$.
\end{enumerate}

\subsection{Threat Model}\label{s:2.5}

We consider the following threat model:
\begin{itemize}
	\item \textit{Adversarial aggregator:}
	Unlike many existing Privacy-Preserving Federated Learning (PPFL) schemes, which assume aggregators are 'honest-but-curious' and will adhere to training protocols for computation, our model permits some adversarial aggregators to participate and collude. This approach is similar to the assumptions in \textsf{TAPFed}. Furthermore, these aggregators may attempt to tamper with the aggregation results, deviate from the correct computation as prescribed by the training rules, to disrupt the normal training process.
	
	\item \textit{Trusted authority:}
	A dependency on a trusted authority is a common feature of most cryptography-based privacy-preserving federated learning schemes, and our solution is no exception. In our framework, this authority handles the configuration of the cryptosystem and the management of keys in training.
\end{itemize}

We assume that the communications between participants occur over secure channels, thereby effectively preventing eavesdropping attacks. Our work focuses on the privacy leakage risks posed by adversarial aggregators and on providing clients with verifiability of the aggregation results to ensure they can obtain the intended training outcomes.
 
	\section{Partial Decryption Verifiable Threshold MCFE}\label{sec:VTMCFE}

\subsection{Definition}\label{s:3.1}

\begin{definition}[Partial Decryption Verifiable Threshold \textsf{MCFE}]
	A $t$-of-$s$ Partial Decryption Verifiable Threshold MCFE (\textsf{VTMCFE}) scheme is comprised of the following algorithms:
\end{definition}

\begin{itemize}
	\item \textsf{Setup} \( (\lambda, t, s, n) \): Takes as input security parameter \( \lambda \), threshold \( t \), total number of clients \( s \), and number of encryption entities \( n \). This algorithm outputs public parameters \( pp \), a master secret key \( msk \), and a set of encryption keys \( \{ek_i\}_{i \in \{1, \dots, n\}} \).
	
	\item \textsf{DKeyGen} \( (pp, msk, \bm{y}) \): Takes as input public parameters \( pp \), master secret key \( msk \), and a function vector \( \bm{y} \). This algorithm generates and outputs a set of functional decryption keys \( \{dk_j\}_{j \in \{1, \dots, s\}} \).
	
	\item \textsf{Encrypt} \( (ek_i, x_i, l) \): Takes as input an encryption key \( ek_i \) (for entity $i$), a message \( x_i \), and a label \( l \). It outputs a ciphertext \( ct_i \).
	
	\item \textsf{Decrypt} \( (pp, \{ct_i\}_{i \in \{1, \dots, n\}}, \bm{y}, dk_j, S) \): Takes as input public parameters \( pp \), a set of ciphertexts \( \{ct_i\}_{i \in \{1, \dots, n\}} \), function vector \( \bm{y} \), functional decryption key \( dk_j \) for entity \( j \), and a subset \( S \) of decryption entities. This algorithm computes and outputs a partial decryption result \( (ct_j', {\Pi}_j ) \), where \( \Pi_j \) is a proof of correctness.
	
	\item \textsf{Verify} \( (pp, \{ct_j', {\Pi}_j\}_{j \in \{1, \dots, s'\}}) \): Takes as input public parameters \( pp \) and a set of \( s' \) partial decryption results \( \{ct_j'\} \) with their corresponding proofs \( \{\Pi_j\} \). This algorithm verifies the correctness of the partial decryptions and outputs a decision (e.g., $1$ for valid, $0$ otherwise).
	
	\item \textsf{CombineRecovery} \( (pp, \{ct_j'\}_{j \in \{1, \dots, s'\}}, l) \): Takes as input public parameters \( pp \), a set of \( s' \) valid partial decryption results \( \{ct_j'\} \) (which have passed verification), and the label \( l \). This algorithm combines these results to recover and output the final function result \( \langle \bm{x}, \bm{y} \rangle \).
\end{itemize}

\subsection{Security Model}\label{s:3.2}

\begin{definition}[IND-Security Game] 
	Adapting the foundational definition from \cite{chotard2018decentralized}, we formalize the IND-Security of our scheme over $n$ senders using the following game, which models the interaction between an adversary $\mathcal{A}$ and a challenger $\mathcal{C}$.
\end{definition}

\textit{$QSetup(\lambda)$:} The challenger $\mathcal{C}$ executes $(pp, msk) \leftarrow \mathsf{Setup}(\lambda)$ to generate the parameters and master secret key. $\mathcal{C}$ also samples a random challenge bit $b \gets \{0,1\}$. The public parameters $pp$ are then sent to the adversary $\mathcal{A}$.

\textit{$QEncrypt(i, x^0, x^1, l)$:} The adversary $\mathcal{A}$ can make unlimited, adaptive queries to a Left-or-Right encryption oracle. Upon querying $(i, x^0, x^1, l)$, $\mathcal{A}$ obtains $ct_{l,i} \leftarrow \mathsf{VTCMFE.Encrypt}(ek_i, x^b, l)$. Subsequent queries for an identical $(l, i)$ pair are disregarded.

\textit{$QDKeyGen(f,j)$:}
$\mathcal{A}$ can adaptively select a function $f$ and query the $\mathsf{VTCMFE.DKeyGen}(msk, f)$. For single function $f$, this query is permitted up to $t-1$ times, with each successful query yielding the functional decryption key $\mathit{dk}_f$. Once the $t-1$ query limit for a specific $f$ is reached, all future queries for that same function $f$ are denied.

\textit{$QCorrupt(i)$:} The adversary $\mathcal{A}$ is permitted to make unlimited and adaptive corruption queries. By querying an index $i$, $\mathcal{A}$ can obtain the corresponding encryption key $ek_i$ for any sender $i$ it selects.

\textit{Finalize:} Ultimately, $\mathcal{A}$ submits its guess $b'$ for the challenge bit $b$. The procedure then concludes by outputting $\beta$, the outcome of the security game, as defined in the subsequent analysis.

The game's final output $\beta$ is determined as follows. Let $CS$ denote the set of corrupted sender indices and let $HS$ be the set of honest (non-corrupted) senders. By default, the output is set to $\beta \leftarrow b'$, which is the adversary's guess.

However, if any of the following three abort conditions occurs, the output is instead set randomly: $\beta \overset{\$}{\leftarrow} \{0, 1\}$.

\begin{enumerate}
	\item An encryption query $QEncrypt(i, x^0_i, x^1_i, l)$ is made for a corrupted sender $i \in CS$, but the challenge plaintexts were different (i.e., $x^0_i \neq x^1_i$).
	
	\item For label $l$, encryption query is issued for at least one honest sender $i \in HS$, but not all honest senders $j \in HS$ were queried for that same label $l$.
	
	\item There exists a label $l$ and a function $f$ (for which $\mathit{dk}_f$ is queried via $QDKeyGen$) that distinguishes the two challenge vectors $x^0 = (x^0_i)_i$ and $x^1 = (x^1_i)_i$. That is, $f(x^0) \neq f(x^1)$, where these vectors are constructed from $QEncrypt$ queries associated with label $l$ as follows:
	\begin{itemize}
		\item For all $i \in CS$ (corrupted senders), the inputs are identical: $x^0_i = x^1_i$.
		\item For all $i \in HS$ (honest senders), the values $x^0_i$ and $x^1_i$ are taken from the corresponding $QEncrypt(i, x^0_i, x^1_i, l)$ queries for that label.
	\end{itemize}
\end{enumerate}

The scheme is considered secure for encrypted data if, for any adversary $\mathcal{A}$, its advantage, defined as
\[
Adv^{IND}_\mathcal{A} = |P[\beta = 1 \mid b = 1] - P[\beta = 1 \mid b = 0]|,
\]
is negligible.

\subsection{Construction}\label{s:3.3}

In this section, we present a detailed description of our partial decryption verifiable threshold MCFE scheme, which consists of six algorithms: \textsf{Setup}, \textsf{DKeyGen}, \textsf{Encrypt}, \textsf{ShareDecrypt}, \textsf{Verify}, and \textsf{CombineRecover}.

\textsf{Setup} $(\lambda, t, s, n)$: Takes as input $\lambda$, threshold \( t \), the number of partial decryption entities \( s \), and the number of encryption entities \( n \), and generates \( \mathcal{G} = (G, p, g, h) \) using \( \mathcal{BG}(1^\lambda) \). Selects hash functions \( \mathcal{H}_1 : \{0, 1\}^* \to G^2 \) , \( \mathcal{H}_2 : \{0, 1\}^* \to \mathbb{Z}_p \). Generates keys \( \bm{s_i} \overset{\$}{\leftarrow} \mathbb{Z}_p^2 \), \( i \in \{1, \dots, n\} \). Randomly chooses\( \{c_{i,k}\}_{i\in \{1,...,t-2\} }\overset{\$}{\leftarrow} \mathbb{Z}_p \) and computes \( \{H_{i,k}\}_{i\in \{2,...,t-1\} } \), where \(H_{i,k}=h^{c_{i-1,k}}\). Defines public parameters \( pp \) as \( (G, p, g, h, \mathcal{H}_1, \mathcal{H}_2, t, s, n, \{H_{i,k}\}_{i\in \{2,...,t-1\} })\), encryption keys \( ek_i = \bm{s_i} \) for \( i \in \{1, \dots, n\} \), and \( msk = \{ ek_i \}_{i \in [n]} \).

\textsf{DKeyGen} $(pp, msk, \bm{y}, k)$: Takes as input $msk$ and a vector \( \bm{y} \) defining an inner product function \( f_y(x) = \langle \bm{x}, \bm{y} \rangle \), and computes function decryption keys \( d = \sum_{i=1}^{n} \bm{s_i} \cdot y_i = (d_1, d_2) \in \mathbb{Z}_p^2 \). Then constructs the following HLR:
\[
\begin{cases}
	w_0 = d_1, w_1 = d_2, w_2 = c_{1,k},\;\dots,\;w_{t-1} = c_{t-2,k};\\[6pt]
	w_{i+t} \;+\; a_1\,w_{i+t-1} \;+\;\dots+\; a_t\,w_i = 0 
	\pmod{p},\quad i \ge 0
\end{cases}
\]
where the auxiliary equation satisfies
\(
(x - \alpha)^t \;=\; x^t \;+\;\sum_{i=1}^t a_i\,x^{\,t-i}
\;\pmod{p}.
\)
Using the above recursion relation, computes \(w_{t+j-1}\) for \(j \in \{1,\dots,s\}\), \(H_{0,k}=h^{d_1} \text{and} H_{1,k}=h^{d_2}\) .
The algorithm outputs the decryption key 
\(
dk_j = w_{\,t+j-1}, j \in \{1,\dots,s\}
\)
for every partial decryption entity and publishes \(H_{0,k},H_{1,k}\).

\textsf{Encrypt} \( (pp, ek_i, x_i, l) \): Takes as input \( pp \), encryption keys \( ek_i \), a message \( x_i \), and a label \( l \). This algorithm computes 
\(
(g^{u_{l,0}}, g^{u_{l,1}}) = \mathcal{H}_1(l) \in G^2,
\)
then outputs the ciphertext as follows:
\[
ct_i = g^{u_l^\top \bm{s_i}} \cdot g^{x_i} = g^{u_l^\top \bm{s_i} + x_i} \in G.
\]

\textsf{ShareDecrypt} \( (pp, \{ct_i\}_{i \in \{1, \dots, n\}}, \bm{y}, dk_j, S, k) \): Takes as input \( \{ ct_i \}_{i \in \{1, \dots, n\}} \), \( pp \), a function \( \bm{y} = \{ y_i \}_{i \in \{1, \dots, n\}} \), and a set of partial decryption entities \( S \). The partial decryption participant \( d_j \in S \) computes the partial decryption result as follows:
\[
ct_j' = (ct_{j,0}', ct_{j,1}', ct_{j,2}'),
\]
where:
\[
ct_{j,0}' = \prod_{i \in \{1, \dots, n\}} ct_i^{y_i},
\]
\[
ct_{j,1}' = g^{u_{l,1} \cdot \frac{w_{t+j-1}}{\alpha^{t+j-1}} \cdot \lambda_{j,1} \cdot \alpha^0},
\]
\[
ct_{j,2}' = g^{u_{l,2} \cdot \frac{w_{t+j-1}}{\alpha^{t+j-1}} \cdot \lambda_{j,2} \cdot \alpha^1}.
\]
Here,\(
\lambda_{j,1} =
\prod_{\substack{j' \in s , j' \neq j}}
\frac{0 - (j' + t - 1)}{j - j'}
\)
\(
\lambda_{j,2} =
\prod_{\substack{j' \in s , j' \neq j}}
\frac{1 - (j' + t - 1)}{j - j'}
\) is the Lagrange basis polynomial. Then it computes the proof for \( ct_{j,1}' \) and \( ct_{j,2}' \) using the DLEQ protocol:
let \(h_{j,1} = g^{u_{l,1} \cdot \lambda_{j,1} \cdot \alpha^{1 - t - j}}\),
\(h_{j,2} = g^{u_{l,2} \cdot \lambda_{j,2} \cdot \alpha^{2 - t - j}}\)

The proof is as follows:
\begin{align*}
	\Pi : \text{Pok}(w_{t+j-1}): \quad & H_{t+j-1} = h^{w_{t+j-1}} \land \\
	& ct_{j,1}' = h_{j,1}^{w_{t+j-1}} \land ct_{j,2}' = h_{j,2}^{w_{t+j-1}}
\end{align*}

The algorithm outputs the partial decryption result \( (ct_j', {\Pi}_j ) \).

\textsf{Verify} \( (pp, \{ct_j'\}_{j \in \{1, \dots, s'\}},k) \): Takes as input \( t \) partial decryption results \( \{ ct_j' \}_{j \in \{1, \dots, s'\}} \). For each \( ct_{j,0}' = \prod_{i=1}^n g^{(u_l^\top \bm{s_i} + x_i) \cdot y_i} \in \{ ct_j' \}_{j \in \{1, \dots, s'\}} \) , the algorithm first verifies if they are equal. Then, the algorithm computes \( \{H_{t+j-1}\}_{j\in\{1,...,s' \}}\),
where \(H_{t+j-1}=H_{t+j-2}^{-a_1}\cdot...\cdot H_{j-1}^{-a_t}=h^{- a_1\,w_{t+j-2}-\dots- a_t\,w_{j-1}} = h^{w_{t+j-1}}\) and using the recursion relation 
.Verifies the proof of \( \{ ct_{j,1}', ct_{j,2}' \}_{j \in \{1, \dots, s'\}} \) using non-interactive DLEQ protocol. If the verification passes, the input \( t \) partial decryption ciphertexts are a valid set of partial decryption ciphertexts.

\textsf{CombineRecover} \( (pp, \{ct_j'\}_{j \in \{1, \dots, s'\}}, l) \): The algorithm takes as input a set of valid partial decryption results \( \{ ct_j' \}_{j \in \{1, \dots, s'\}} \), and outputs the recovery result:
\[
[\beta] = \frac{ct_{j,0}'}{\left( \prod_{j \in \{1, \dots, s'\}} ct_{j,1}' \right) \cdot \left( \prod_{j \in \{1, \dots, s'\}} ct_{j,2}' \right)},
\]
and ultimately solves the discrete logarithm problem to obtain the function value \(\beta = \langle \bm{x}, \bm{y} \rangle\).

\textit{Correctness. }
\begin{align*}
	[\beta] &= \frac{ct_{j,0}'}{\left( \prod_{j \in \{1, \dots, s'\}} ct_{j,1}' \right) \cdot \left( \prod_{j \in \{1, \dots, s'\}} ct_{j,2}' \right)} \\
	&= \frac{g^{\sum_{i=1}^n \bm{u_l}^\top \bm{s_i} y_i + \sum_{i=1}^n x_i y_i}}{
		g^{u_{l,1} \cdot \alpha^0 \cdot \sum_{j} \frac{w_{t+j-1}}{\alpha^{t+j-1}} \lambda_{j,1}} 
		\cdot 
		g^{u_{l,2} \cdot \alpha^1 \cdot \sum_{j} \frac{w_{t+j-1}}{\alpha^{t+j-1}} \lambda_{j,2}}
	} \\
	&= \frac{g^{\sum_{i=1}^n \bm{u_l}^\top \bm{s_i} y_i + \sum_{i=1}^n x_i y_i}}{
		g^{u_{l,1} \cdot \alpha^0 \cdot p(0)} 
		\cdot 
		g^{u_{l,2} \cdot \alpha^1 \cdot p(1)}
	} \\
	&= \frac{g^{\sum_{i=1}^n \bm{u_l}^\top \bm{s_i} y_i + \sum_{i=1}^n x_i y_i}}{
		g^{u_{l,1} \cdot w_0} 
		\cdot 
		g^{u_{l,2} \cdot w_1}
	} \\
	&= \frac{g^{\sum_{i=1}^n \bm{u_l}^\top \bm{s_i} y_i + \sum_{i=1}^n x_i y_i}}{
		g^{\bm{u_l}^\top \bm{d}}
	} \\
	&= \frac{g^{\sum_{i=1}^n \bm{u_l}^\top \bm{s_i} y_i + \sum_{i=1}^n x_i y_i}}{
		g^{\bm{u_l}^\top \sum_{i=1}^n y_i \bm{s_i}}
	} = g^{\sum_{i=1}^n x_i y_i}.
\end{align*}

\subsection{Security Analysis}\label{s:3.4}

In order to prove the security of our scheme, we designed a series of hybrid games. Under the DDH assumption, the real security game is gradually transformed into an ideal game in which the opponent's advantage can be ignored.

\begin{theorem}[\textsf{IND}-Security]\label{thm1}
	We assert that the presented \textsf{MCFE} protocol achieves \textsf{IND}-security within the random oracle model, contingent upon the \textsf{DDH} assumption. Specifically, for a PPT adversary $\mathcal{A}$, the advantage is bounded as follows:
	\[
	\mathsf{Adv}_{\mathcal{G}, \mathcal{A}}^{\textsf{IND}} \leq 2M \cdot \mathsf{Adv}_{\mathcal{G}}^{\textsf{ddh}}(t) + \mathsf{Adv}_{\mathcal{G}}^{\textsf{ddh}}(t + 4M \times t_{\mathbb{G}}) + \frac{2M}{p},
	\]
	where $M$ represents the queries number made to the random oracle $\mathcal{H}$, and $t_{\mathbb{G}}$ denotes the computational cost of a single exponentiation in the group $\mathbb{G}$.
\end{theorem}

\begin{IEEEproof}
	As shown in Fig.~\ref{fig:game1}, our proof is carried out through mixed games. We suppose that $\mathcal{A}$ is a probability polynomial time (\textsf{PPT}) adversary. For each game $G_{\text{index}}$ in the sequence, we define the adversary's advantage as $\mathsf{Adv}_{\text{index}} = \left|\Pr[G_{\text{index}}(\mathcal{A})|b = 1] - \Pr[G_{\text{index}}(\mathcal{A})|b = 0]\right|$, with the probability space defined by the random coins of both the game $G_{\text{index}}$ and the adversary $\mathcal{A}$. Furthermore, we use the notation $G_{\text{index}}(\mathcal{A})$ (or simply $G_{\text{index}}$ when the context is clear) to represent the event where the \textsf{Finalize} procedure (as per Definition~5) outputs $\beta = 1$, based on the adversary's final guess $b'$ during the interaction.
	
	\begin{figure*}[t]
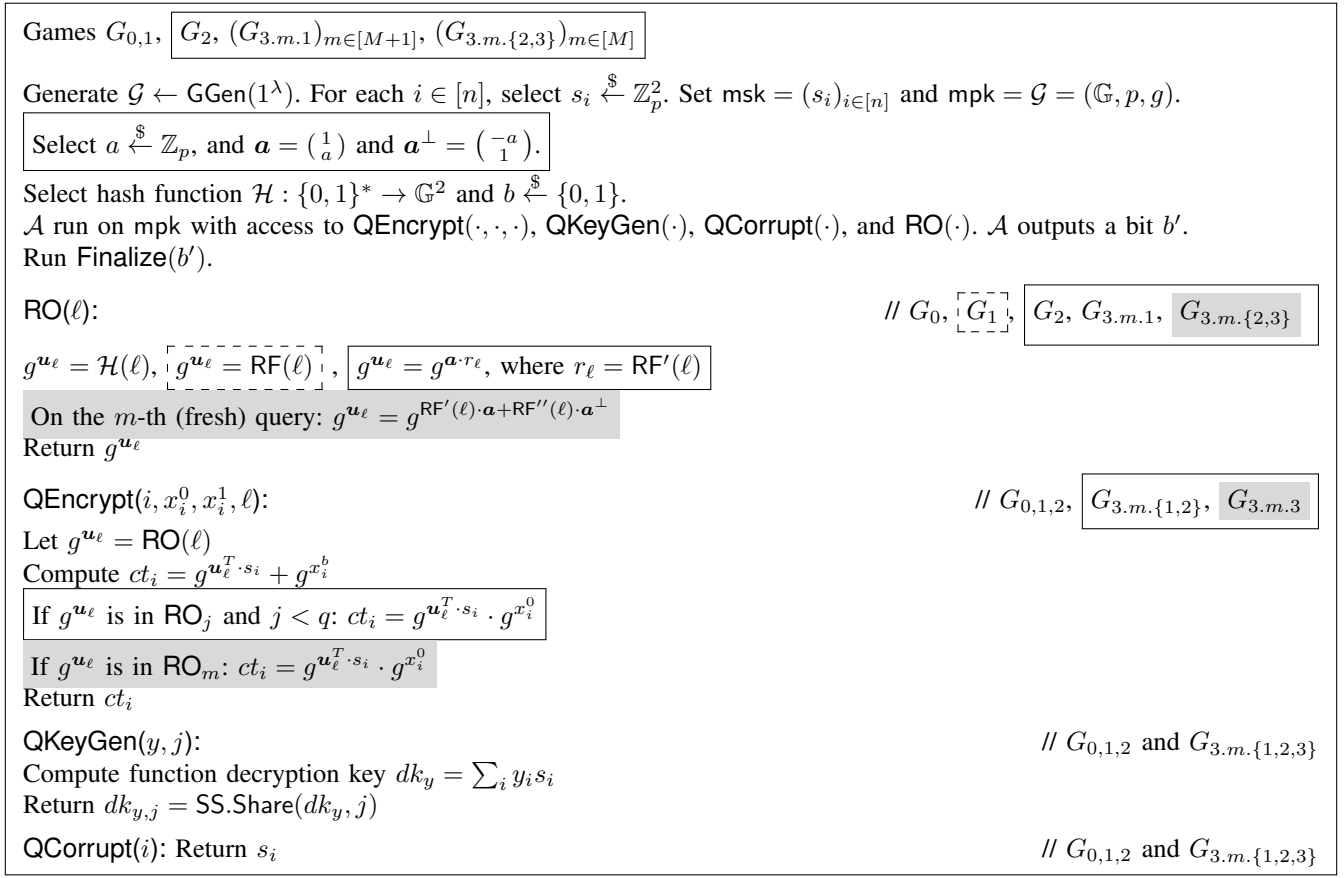

		\centering
		\fbox{
			\begin{minipage}{0.95\linewidth}
				Games $G_{0,1}$, \fbox{{$G_2$}, $(G_{3. m. 1})_{m \in [M+1]}$, $(G_{3. m. \{2,3\}})_{m \in [M]}$}
				
				\vspace{1ex}

				Generate $\mathcal{G} \leftarrow \mathsf{GGen}(1^\lambda)$. For each $i \in [n]$, select $s_i \overset{\$}{\leftarrow} \mathbb{Z}_p^2$. Set $\mathsf{msk} = (s_i)_{i \in [n]}$ and $\mathsf{mpk} = \mathcal{G} = (\mathbb{G}, p, g)$.
					
				\fbox{Select $a \overset{\$}{\leftarrow} \mathbb{Z}_p$, and $\bm{a} = \left(\begin{smallmatrix} 1 \\ a \end{smallmatrix}\right)$ and $\bm{a}^\perp = \left(\begin{smallmatrix} -a \\ 1 \end{smallmatrix}\right)$.}
					
				Select hash function $\mathcal{H}: \{0,1\}^* \to \mathbb{G}^2$ and $b \overset{\$}{\leftarrow} \{0,1\}$.
					
				$\mathcal{A}$ run on $\mathsf{mpk}$ with access to $\textsf{QEncrypt}(\cdot,\cdot,\cdot)$, $\textsf{QKeyGen}(\cdot)$, $\textsf{QCorrupt}(\cdot)$, and $\textsf{RO}(\cdot)$. $\mathcal{A}$ outputs a bit $b'$.
					
				Run $\textsf{Finalize}(b')$.

				\vspace{1ex}
				\textsf{RO($\ell$):} \hfill // $G_0$, \dbox{$G_1$}, \fbox{$G_2$, $G_{3.m.1}$, \colorbox{gray!30}{$G_{3. m. \{2,3\}}$} } 
				
				$g^{\bm{u}_\ell} = \mathcal{H}(\ell)$, \dbox{{$g^{\bm{u}_\ell} = \mathsf{RF}(\ell)$}} , \fbox{$g^{\bm{u}_\ell} = g^{\bm{a} \cdot r_\ell}$, where $r_\ell = \mathsf{RF'}(\ell)$} 
				
				\colorbox{gray!30}{On the $m$-th (fresh) query: $g^{\bm{u}_\ell} = g^{\mathsf{RF'}(\ell) \cdot \bm{a} + \mathsf{RF''}(\ell) \cdot \bm{a}^\perp}$} 
				
				Return $g^{\bm{u}_\ell}$
				
				\vspace{1ex}
				\textsf{QEncrypt($i, x_i^0, x_i^1, \ell$):} \hfill // $G_{0,1,2}$, \fbox{$G_{3. m. \{1,2\}}$, \colorbox{gray!30}{$G_{3.m.3}$}} 
				
				Let $g^{\bm{u}_\ell} = \textsf{RO}(\ell)$
				
				Compute $ct_i = g^{\bm{u}_\ell^T \cdot s_i} + g^{x_i^b}$
				
				\fbox{If $g^{\bm{u}_\ell}$ is in $\textsf{RO}_j$ and $j < q$:
					$ct_i = g^{\bm{u}_\ell^T \cdot s_i} \cdot g^{x_i^0}$} 
				
				\colorbox{gray!30}{If $g^{\bm{u}_\ell}$ is in $\textsf{RO}_m$:
					$ct_i = g^{\bm{u}_\ell^T \cdot s_i} \cdot g^{x_i^0}$}

				Return $ct_i$
				
				\vspace{1ex}
				\textsf{QKeyGen($y,j$):} \hfill // $G_{0,1,2}$ and $G_{3. m. \{1,2,3\}}$
				
				Compute function decryption key $dk_y = \sum_i y_i s_i$
				
				Return $dk_{y,j} = \mathsf{SS.Share}(dk_y,j)$
				
				\vspace{1ex}
				\textsf{QCorrupt($i$):} Return $s_i$ \hfill // $G_{0,1,2}$ and $G_{3. m. \{1,2,3\}}$
				
			\end{minipage}
		}
		\caption{Games $G_{0,1,2}$ and $G_{3. m. \{1,2,3\}}$ for the proof.}
		\label{fig:game1}
	\end{figure*}

	\vspace{1ex}
	
	\paragraph{Game $G_0$}
	This initial game, $G_0$, corresponds to \textsf{IND}-security experiment outlined in Definition~2. In this game, hash function $\mathcal{H}$ is defined as a random oracle (\textsf{RO}), and its output is mapped to $\mathbb{G}^2$. Its primary role is to generate $g^{\bm{u}_\ell}$ by computing $\mathcal{H}(\ell)$.
	
	\paragraph{Game $G_1$}
	In $G_1$, we modify the simulation. Answers to all new queries to the random oracle are now generated as uniformly random pairs from $\mathbb{G}^2$. This change is purely conceptual, as the oracle was already random; thus, the simulation is perfect. Consequently, the adversary's advantage remains unchanged: $\mathsf{Adv}_0 = \mathsf{Adv}_1$.
	
	\paragraph{Game $G_2$}
	We introduce another modification. \textsf{RO}-queries are now answered with a random pair drawn from the span of $g^{\bm{a}}$, where $\bm{a}$ is defined as $\left(\begin{smallmatrix} 1 \\ a \end{smallmatrix}\right)$ for a randomly chosen $a \leftarrow \mathbb{Z}_p$. This transition depends on the Multi-DDH problem. The difference in advantage is bounded by: $\mathsf{Adv}_1 - \mathsf{Adv}_2 \leq \mathsf{Adv}_{\mathcal{G}}^{ddh}(t + 4M \times t_{\mathbb{G}})$, where $M$ represents again \textsf{RO}-queries $t_{\mathbb{G}}$ is the cost of an exponential operation.
	
	\paragraph{Game $G_3$}
	We alter \textsf{QEncrypt} simulation. All encryption queries are now processed using the message $x_i^0$, regardless of the challenge $b$. Concurrently, we revert simulation of the \textsf{RO} queries back to $G_1$'s method, answering them with uniformly random pairs from $\mathbb{G}^2$.

It is evident that in this final game, the adversary's advantage becomes exactly 0, as the challenge bit $b$ is completely absent from the simulation. We use a hybrid argument applied sequentially to the \textsf{RO}-queries to trans from $G_2$ to $G_3$. We introduce a series of games indexed by $m$, for $m = 1, \ldots, M$. Note that this counts only the distinct queries to the \textsf{RO}, as any repeated query receives its original answer. We provide this proof in detail due to the importance of the technique.

$G_{3.1.1}:$
This game is defined to be identical to $G_2$. Consequently, their advantages are equal: $\mathsf{Adv}_{G_2} = \mathsf{Adv}_{G_{3.1.1}}$.

$G_{3.m.1} \rightsquigarrow G_{3.m.2}:$
For the $m$-th \textsf{RO}-query, we first alter the distribution of its response. Leveraging the \textsf{DDH} assumption, the output distribution is modified, transitioning from a uniform distribution over the subspace spanned by $g^{\bm{a}}$ to a uniform distribution across the entire group $\mathbb{G}^2$.

To analyze this change, we express a uniformly random vector from $\mathbb{Z}_p^2$ using the basis $\left(\left(\begin{smallmatrix} 1 \\ a \end{smallmatrix}\right), \left(\begin{smallmatrix} -a \\ 1 \end{smallmatrix}\right)\right)$. A vector can thus be written as $u_1 \cdot \bm{a} + u_2 \cdot \bm{a}^\perp$, where both $u_1, u_2$ are randomly sampled from $\mathbb{Z}_p$.

Finally, we transition to a slightly different distribution, $u_1 \cdot \bm{a} + u_2 \cdot \bm{a}^\perp$, where $u_1$ is still drawn from $\mathbb{Z}_p$, but $u_2$ is drawn from $\mathbb{Z}_p^*$ (i.e., $u_2 \ne 0$). The maximum impact of this final change on the adversary's view is a distance of $1/p$ in static. This leads to the bound: $\mathsf{Adv}_{3.m.1} - \mathsf{Adv}_{3.m.2} \leq \mathsf{Adv}_{\mathcal{G}}^{\textsf{ddh}}(t) + 1/p$. The requirement $u_2 \in \mathbb{Z}_p^*$ is a crucial detail needed to ensure that $\bm{u}_\ell^{\top} \cdot \bm{a}^\perp \ne 0$.

\begin{figure*}[t] 
	\centering
	\fbox{
		\begin{minipage}{0.95\linewidth}
			Games $(G_{3.m.2}^*, G_{3.m.3}^*)_{m \in [M]}$:
			
			\vspace{1ex}
			
			$(\textsf{state}, (z_i \in \mathbb{Z}_p^2 \cup \{\perp\})_{i \in [n]}) \leftarrow \mathcal{A}(1^\lambda, 1^n)$
			
			Generate $\mathcal{G} \leftarrow \mathsf{GGen}(1^\lambda)$. For each $i \in [n]$, sample $s_i \overset{\$}{\leftarrow} \mathbb{Z}_p^2$. Set $\mathsf{msk} = (s_i)_{i \in [n]}$ and $\mathsf{mpk} = \mathcal{G} = (\mathbb{G}, p, g)$.
			
			Sample $a \overset{\$}{\leftarrow} \mathbb{Z}_p$, and define $\bm{a} = \left(\begin{smallmatrix} 1 \\ a \end{smallmatrix}\right)$ and $\bm{a}^\perp = \left(\begin{smallmatrix} -a \\ 1 \end{smallmatrix}\right)$.
			
			Sample a bit $b \overset{\$}{\leftarrow} \{0,1\}$
			
			The adversary $\mathcal{A}$ is run on $\mathsf{mpk}$ with access to oracles $\textsf{QEncrypt}(\cdot,\cdot,\cdot)$, $\textsf{QKeyGen}(\cdot)$, $\textsf{QCorrupt}(\cdot)$, and $\textsf{RO}(\cdot)$. $\mathcal{A}$ outputs a bit $b'$.
			
			Run $\textsf{Finalize}(b')$.
			
			\vspace{1ex}
			\textsf{RO($\ell$):} \hfill // $G_{3.m.\{2,3\}}^*$
			
			$g^{\bm{u}_\ell} = g^{\bm{a} \cdot r_\ell}$, where $r_\ell = \mathsf{RF'}(\ell)$
			
			On the $m$-th (fresh) query: $g^{\bm{u}_\ell} = g^{\mathsf{RF'}(\ell) \cdot \bm{a} + \mathsf{RF''}(\ell) \cdot \bm{a}^\perp}$
			
			Return $g^{\bm{u}_\ell}$

			\vspace{1ex}
			\textsf{QEncrypt($i, x_i^0, x_i^1, \ell$):} \hfill // \fbox{$G_{3.m.2}^*$}, \colorbox{gray!30}{$G_{3.m.3}^*$}
			
			Let $g^{\bm{u}_\ell} = \textsf{RO}(\ell)$
			
			Compute $ct_i = g^{{\bm{u}_\ell}^{\top}\cdot s_i} \cdot g^{x_i^b}$
			
			If $g^{\bm{u}_\ell}$ is in $\textsf{RO}_j$ and $j < q$: $ct_i = g^{{\bm{u}_\ell}^{\top}\cdot s_i} \cdot g^{x_i^b}$
			
			If $g^{\bm{u}_\ell}$ is in $\textsf{RO}_m$:
			
			\hspace*{2ex}• If $(x_i^0, x_i^1) \ne z_i$, the game ends and returns $\beta \overset{\$}{\leftarrow} \{0,1\}$
			
			\hspace*{2ex}• Otherwise, set $ct_i = g^{{\bm{u}_\ell}^{\top}\cdot s_i}  \fbox{$ \cdot g^{x_i^b}$}\colorbox{gray!30}{$ \cdot g^{x_i^0}$}$, $S = S \cup \{i\}$
			
			Return $ct_i$
			
			\vspace{1ex}
			\textsf{QKeyGen($y,j$):} \hfill // $G_{3.m.\{2,3\}}^*$
			
			Compute function decryption key $dk_y = \sum_i y_i s_i$
			
			Return $dk_{y,j} = \mathsf{SS.Share}(dk_y,j)$
			
			\vspace{1ex}
			\textsf{QCorrupt($i$):} \hfill // $G_{3.m.\{2,3\}}^*$
			
			If $z_i = (x_i^0, x_i^1)$ with $x_i^0 \ne x_i^1$, the game ends and returns $\beta \overset{\$}{\leftarrow} \{0,1\}$
			
			Return $s_i$
		\end{minipage}
	}
	\caption{Games $G_{3.m.\{2,3\}}^*$, $m \in [M]$ for the proof.}
\end{figure*}

$G_{3.m.2} \rightsquigarrow G_{3.m.3}:$
The ciphertext generation process is modified from $ct_i = g^{{\bm{u}_\ell}^{\top}\cdot s_i} \cdot g^{x_i^b}$ to $ct_i = g^{{\bm{u}_\ell}^{\top}\cdot s_i} \cdot g^{x_i^0}$, where $g^{\bm{u}_\ell}$ represents the $m$-th \textsf{RO}-query. It will now be demonstrated that this modification is imperceptible to the adversary.

When there is none \textsf{QEncrypt}-queries use $\textsf{RO}_m$, games $G_{3.m.\{2,3\}}$ are trivially identical. But it shows that $\textsf{Adv}_{3.m.2} = \textsf{Adv}_{3.m.3}$ holds even in the case where this \textsf{RO}-query output is indeed utilized by \textsf{QEncrypt}-queries. We establish this assertion using a two-step argument.

\textit{Step 1.} First, we demonstrate a PPT adversary $\mathcal{B}^*$ that plays against selective variants of the games, $G_{3.m.2}^*$ and $G_{3.m.3}^*$ (defined in Fig.~4). In these selective games, \textsf{QCorrupt} queries must be issued before the initialization phase. We show that the advantages are related as follows for $t=2,3$:
\[
\textsf{Adv}_{3.m.t} = (p^2 + 1)^n \cdot \textsf{Adv}_{3.m.t}^*(\mathcal{B}^*).
\]

\textit{Step 2.} Second, we prove that for any PPT adversary $\mathcal{B}^*$, its advantages in these selective games are equal: $\textsf{Adv}_{3.m.2}^*(\mathcal{B}^*) = \textsf{Adv}_{3.m.3}^*(\mathcal{B}^*)$. The completion of these two steps concludes the proof.

\textit{Details of Step 1.}
We construct the adversary $\mathcal{B}^*$ to play against $G_{3.m.t}^*$ ($t=2,3$) such that the aforementioned advantage relationship holds.

$\mathcal{B}^*$ initiates by guessing, for all $i \in [n]$, a value $z_i \leftarrow \mathbb{Z}_p^2 \cup \{\perp\}$. It sends this complete set of guesses to its selective game $G_{3,m,t}^*$. Each $z_i$ represents $\mathcal{B}^*$'s guess for the \textsf{QEncrypt} query:
either a pair $(x_i^0, x_i^1)$ or $\perp$ (signifying no query for index $i$).
$\mathcal{B}^*$ then simulates the view for the original adversary $\mathcal{A}$ using its own oracles.

Let $E$ be the event that all of $\mathcal{B}^*$'s guesses are correct.
If $E$ occurs, $\mathcal{B}^*$ perfectly simulates $\mathcal{A}$'s view as in $G_{3.m.t}$.
If $E$ does not occur (the guess was incorrect), $\mathcal{B}^*$ aborts the simulation and outputs a random bit $\beta$.

The probability of this success event $E$ is $P_E = (p^2 + 1)^{-n}$, and $E$ is independent of $\mathcal{A}$'s view. We define $P(b, E) = \Pr[G_{3.m.t}|b=b, E]$. The advantage of $\mathcal{B}^*$ in its selective game, $\textsf{Adv}_{G_{3.m.t}^*}(\mathcal{B}^*)$, is calculated as:
\begin{align*}
	&\left| \left( P(0, E)\cdot P_E + \frac{1-P_E}{2} \right) - \left( P(1, E)\cdot P_E + \frac{1-P_E}{2} \right) \right| \\
	&= \left| P(0, E)\cdot P_E - P(1, E)\cdot P_E \right| \\
	&= P_E\cdot \left| P(0, E) - P(1, E) \right| \\
	&= P_E\cdot  \textsf{Adv}_{3.m.t} \\
	&= (p^2 + 1)^{-n} \cdot \textsf{Adv}_{3.m.t}.
\end{align*}
Rearranging this equation gives the relation claimed in Step 1.

\textit{Details of Step 2.}
Conditioned on event $E'$ (consistent guesses) occurring, we will now demonstrate that the distributions of games $G_{3.m.2}^*$ and $G_{3.m.3}^*$ are identical. This proof relies on the statistical indistinguishability of the following two distributions, which holds for any choice of $\gamma$:
\begin{gather*}
	(s_i)_{\substack{i \in [n] \\ z_i = (x_i^0, x_i^1)}} 
	\quad \text{and} \quad
	\left(s_i + \bm{a}^\perp \cdot \gamma (x_i^b - x_i^0)\right)_{\substack{i \in [n] \\ z_i = (x_i^0, x_i^1)}}
\end{gather*}

Here, $\bm{a}^\perp$ is defined as $\left( \begin{smallmatrix} -a \\ 1 \end{smallmatrix} \right) \in \mathbb{Z}_p^2$, and each $s_i$ is drawn uniformly at random from $\mathbb{Z}_p^2$. The identical nature of the distributions holds because the $s_i$ values are independent of the $z_i$ guesses. It is crucial to note that this independence is a consequence of the selective security setting; it would not be guaranteed with adaptive \textsf{QEncrypt}-queries. Consequently, we can substitute $s_i$ with $s_i + \bm{a}^\perp \cdot \gamma (x_i^b - x_i^0)$ without altering the game's overall distribution.

We trace where the term $\bm{a}^\perp \cdot \gamma (x_i^b - x_i^0)$ impacts the adversary's view:

\begin{itemize}
	\item \textsf{QCorrupt}: The term has no effect. We are conditioned on $E'$, which guarantees that for any $i$ with $z_i \neq \perp$, $i$ is not corrupted or $x_i^1 = x_i^0$.
	
	\item \textsf{QKeyGen}($y$): The term could potentially modify the key to:
	\[
	\mathsf{dk}_y = \sum_{i \in [n]} s_i \cdot y_i + \bm{a}^\perp \cdot \gamma \sum_{i \in [n], z_i = (x_i^0, x_i^1)} y_i (x_i^b - x_i^0).
	\]
	This additional component vanishes. The conditions of $E'$ ensure the summation $\sum_{i : z_i = (x_i^0, x_i^1)} y_i (x_i^b - x_i^0)$ is always zero, as $z_i \neq \perp$ implies $x_i^1 = x_i^0$ for all relevant indices $i$.
	
	\item \textsf{QEncrypt}: The term appears only in \textsf{QEncrypt}-queries using the $m$-th \textsf{RO}-query, $g^{\bm{u}_\ell}$. For all other \textsf{RO}-queries, $g^{\bm{u}_\ell}$ is in the span of $[a]$, and $\bm{a}^\top \bm{a}^{\perp} = 0$. For the $m$-th query, the ciphertext becomes:
	\[
	ct_i = g^{\bm{u}_\ell^\top \cdot s_i} \cdot g^{\bm{u}_\ell^\top \cdot \bm{a}^{\perp} \gamma(x_i^b - x_i^0)} \cdot g^{x^b_i} .
	\]
	Since $\bm{u}_\ell^\top \bm{a}^\perp \neq 0$ (by setup), we set $\gamma \leftarrow -1/ (\bm{u}_\ell^\top \bm{a}^\perp) \mod p$. This constant $\gamma$ is independent of $i$ and transforms the ciphertext to $ct_i = g^{{\bm{u}_\ell^\top}\cdot s_i} \cdot g^{x^0_i}$, simultaneously changing all $x_i^b$ encryptions to $x_i^0$.
\end{itemize}

By reversing this statistically perfect substitution, we confirm that the original $ct_i$ (in $G_{3.m.2}^*$) is identically distributed to $g^{\bm{u}_\ell^\top \cdot s_i} \cdot g^{x_i^0}$, which is precisely the form of the ciphertext in game $G_{3.m.3}^*$.

The games are identical conditioned on $E'$ (consistent guesses). If $\neg E$ (incorrect guess) occurs, $\mathcal{B}^*$ aborts with a random bit in both games. Therefore, $\textsf{Adv}_{3.m.2}^*(\mathcal{B}^*) = \textsf{Adv}_{3.m.3}^*(\mathcal{B}^*)$, implying $\textsf{Adv}_{3.m.2} = \textsf{Adv}_{3.m.3}$.

$G_{3.m.3} \rightsquigarrow G_{3.m+1.1}: $ This step reverses the $G_{3.m.1} \rightsquigarrow G_{3.m.2}$ transition. By the \textsf{DDH} assumption, we revert the $m$-th \textsf{RO}-query $g^{\bm{u}_\ell}$ from uniformly random in $\mathbb{G}^2$ (where $\bm{u}_\ell^\top \bm{a}^\perp \neq 0$) to uniformly random in the span of $\bm{a}$.
\[
\textsf{Adv}_{3.m.3} - \textsf{Adv}_{3.m+1.1} \leq \textsf{Adv}_{\mathcal{G}}^{\textsf{ddh}}(t) + 1/p.
\]

In summary, by iterating this process through all $M$ \textsf{RO}-queries, we find that $G_{3.M+1.1}$ is equivalent to $G_3$. This yields the bound:
\[
\textsf{Adv}_2 - \textsf{Adv}_3 \leq 2M (\textsf{Adv}_{\mathcal{G}}^{\textsf{ddh}}(t) + 1/p).
\]
Furthermore, since the advantage in game $G_3$ is zero ($\textsf{Adv}_3 = 0$), the proof is complete.
\end{IEEEproof}

	\section{Federated Learning Secure Aggregation}\label{sec:VTSAFL}

In this section, we present \textsf{VTSAFL}, a new verifiable threshold secure aggregation protocol for privacy-preserving federated learning, which is based on our novel verifiable threshold \textsf{MCFE} scheme. We will detail the system framework, which consists of clients, distributed aggregators, and a trusted authority (TA), as well as the distributed training process. A security analysis will follow, examining privacy risks and attacks from malicious aggregators.

\begin{figure}[!h]
	\centerline{\includegraphics[width=0.5\textwidth]{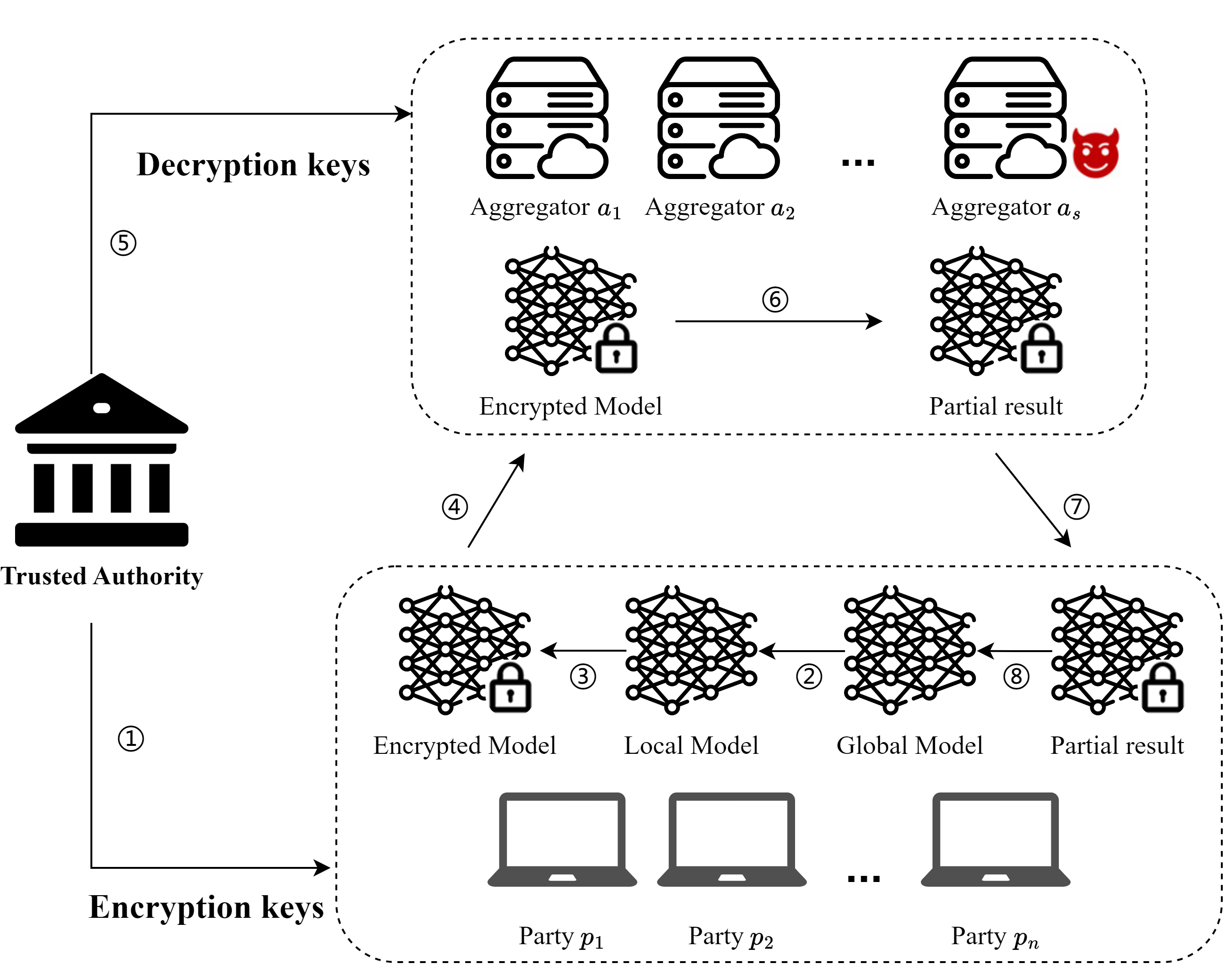}}
	
	\caption{Our Privacy-Preserving Federated Learning Framework }
	\label{fig:PPFLframework}
	
\end{figure}

\subsection{System Architecture}

The \textsf{VTSAFL} framework, illustrated in Fig.~\ref{fig:PPFLframework}, consists of three primary entities. A trusted authority (TA) initializes the cryptosystem and distributes keys. A set of clients uses local data to train and subsequently encrypt their models. Finally, a group of distributed aggregators collects these encrypted models, decrypts them, and computes an accompanying proof of decryption. This allows all clients to verify the correctness of the aggregation result.

It is important to note that, unlike previous solutions based on functional encryption, we assume a more robust threat model wherein aggregators can be malicious. These aggregators may collude to breach user privacy or alter results to disrupt training, an assumption that aligns with the \textsf{TAPFed} framework \cite{TAPFed}.

By virtue of the system's threshold nature, participation from the entire set of aggregators is not necessary in each round. The aggregators do not communicate with or even need to be aware of each other. This feature significantly enhances both the stability and the operational efficiency of the system.

\begin{algorithm}[!ht]
	\caption{Federated Learning Framework}
	\label{alg:fl_framework}
	\SetKwInOut{Input}{Input}
	\SetKwInOut{Output}{Output}
	
	\Input{
		The security parameter $\lambda$, party set $S_P$ and each party $p_i \in S_P$ has its dataset $D_i$,
		aggregator set $S_A$, trusted authority $T$,
		maximum training round $K$, hyperparameters $hp$,
		encryption label for each training round $\{l^{(k)}\}_{k \in \{1,...,K\}}$
	}
	\Output{Global model $\theta_G$}
	
	\BlankLine
	\textbf{Initialize:} \\
	$T$ runs $(pp, msk) \leftarrow \mathsf{VTMCFE.Setup}(\lambda, t, s, n)$ and broadcasts $pp$\; 
	$T$ sends $ek_i$ to $p_i$ for $i \in \{1,...,n\}$\;

	\textbf{Training:} \\
	\SetKwRepeat{Do}{do}{while}
	\Do{$k \leq K$}{
		\textbf{Local model training:}\\
		\textbf{if} $k == 1$ \textbf{then} $p_i$ initializes local model $\theta_{p_i}^{(0)}$\;
		$\theta_{p_i}^{(k)} \leftarrow p_i$ trains model with $(D_i, \theta_{p_i}^{(0)}, hp)$\;
		$p_i$ runs $ct^{(k)}_{p_i} \leftarrow \mathsf{VTMCFE.Encrypt}(ek_i,\theta_{p_i}^{(k)},l^{(k)})$\;
		$p_i$ sends $ct^{(k)}_{p_i}$ to $a_j \in S_A$\;
		\textbf{FedAvg model aggreation:}\\
		$a_j$ obtains all $ct^{(k)}_{p_i}$ from $p_i \in S_P$\;
		$a_j$ prepares fusion weight $y^{a_j}$ based on $\{ct^{(k)}_{p_i}\}$\;
		$a_j$ requests $dk_j$ from $T$ with $(y^{a_j})$\;
		$T$ checks and validates all requests collected from $a_j \in S_A$\; 
		$T$ chooses a valid request $y$ and runs $\{dk_j\} \leftarrow \mathsf{VTMCFE.DKeyGen}(pp, msk, y)$\;
		$T$ sends $dk_j$ to $a_j \in S_A$\;
		$a_j$ runs $(ct^{'(k)}_j,\Pi_j) \leftarrow \mathsf{VTMCFE.Decrypt}(pp,\{ct^{(k)}_{p_i}\}, y, dk_j)$\;
		$a_j$ sends $(ct^{'(k)}_j,\Pi_j)$ to $p_i \in S_P$\;
		\textbf{Global model updating:}\\
		$p_i$ runs $\mathsf{VTMCFE.Verify}(pp, \{ct_j', {\Pi}_j\}_{j \in \{1, \dots, s'\}})$\; 
		\If{$\mathsf{VTMCFE.Verify}(pp, \{ct_j', {\Pi}_j\}_{j \in \{1, \dots, s'\}}) == 1$}{
			$p_i$ runs $\theta_{A}^{(k)} \leftarrow \mathsf{VTMCFE.CombineRecovery}(pp, \{ct_j'\}_{j \in \{1, \dots, s'\}}, l) $\;
			$p_i$ updates $\theta_{G}^{(k)} \leftarrow \theta_{A}^{(k)}$\;
		}
		
	}
\end{algorithm}

\subsection{Training Process} 
The distributed training process of the \textsf{VTSAFL} framework is based on the traditional federated learning workflow and is primarily divided into five stages: initialization, local model training and encryption, aggregation and partial decryption by aggregators, verification of decrypted results by clients, and finally, global model updates. The entire process is detailed in Algorithm \ref{alg:fl_framework}.

\textit{Initialization}: The Trusted Authority (TA) runs the \texttt{\textsf{VTMCFE.Setup}} algorithm to generate public parameters \texttt{pp} and a master secret key \texttt{msk}. The public parameters are broadcast to all participants, while the TA securely sends the respective encryption keys \texttt{$ek_i$} to each client \texttt{$p_i$}.

\textit{Local Model Training and Encryption}: Each client \texttt{$p_i$} first initializes its local model. Then, for each training round, it trains its local model using its own private dataset \texttt{$D_i$}. After training, the client encrypts its updated local model $\theta_{p_i}^{(k)}$ using its encryption key and the current round's label $l^{(k)}$, producing a ciphertext $ct_{p_i}^{(k)}$. This encrypted model is then sent to the aggregators.

\textit{Aggregation and Partial Decryption}: After receiving the encrypted models from all clients, each aggregator \texttt{$a_j$} prepares a fusion weight vector \texttt{y} and requests the corresponding decryption key share from the TA. The TA validates these requests, generates the decryption key shares $\{dk_j\}$ using \texttt{\textsf{VTMCFE.DKeyGen}}, and distributes them to the respective aggregators. Each aggregator \texttt{$a_j$} then uses its key share \texttt{$dk_j$} to compute a partial decryption result of the aggregated model and a corresponding proof of correctness $(ct_j^{(k)}, \Pi_j)$ by running the \texttt{\textsf{VTMCFE.Decrypt}} algorithm. This partial result and proof are then sent back to the clients.

\textit{Verification}: Upon receiving the partial results and proofs from a threshold number of aggregators, each client \texttt{$p_i$} runs the \texttt{\textsf{VTMCFE.Verify}} algorithm. This step allows clients to check the validity of the computation performed by each aggregator, thus protecting against malicious aggregators who might return incorrect results to disrupt the training process.

\textit{Global Model Update}: If the verification is successful, the clients proceed to the final step. They run the \texttt{\textsf{VTMCFE.CombineRecovery}} algorithm to combine the valid partial results and recover the final aggregated global model $\theta_G^{(k)}$. This new global model is then used as the starting point for the next round of local training. This iterative process continues for a predetermined number of rounds $K$.

\subsection{Security of VTSAFL}
In this section, we analyze the security of our \textsf{VTSAFL} framework. Our threat model considers the aggregators to be adversarial, meaning they may maliciously deviate from the protocol to compromise the system's integrity or privacy. We assume that up to $t-1$ aggregators can collude. Communications are secured, and the Trusted Authority (TA) is considered trustworthy. The \textsf{VTSAFL} framework provides resilience against several privacy and integrity attacks as follows.

(i) Gradient Inference and Intermediate Model Leakage:
A primary privacy risk in FL is that adversaries could infer sensitive information from shared model updates. In our \textsf{VTSAFL} framework, clients encrypt their local models before uploading them. Consequently, an aggregator cannot directly access any individual client's model. Furthermore, because we employ a $(t, s)$-threshold architecture, no single aggregator possesses the complete functional key. An intermediate model can only be decrypted with the cooperation of at least $t$ aggregators. Therefore, any collusion of up to $t-1$ adversarial aggregators is insufficient to decrypt the aggregated intermediate model, thus effectively preventing intermediate model leakage.

(ii) Collusion Attack:
Our framework is designed to be resilient against collusion attacks involving up to $t-1$ aggregators. The functional decryption key is distributed among $s$ aggregators using a secret sharing scheme, where a threshold of $t$ shares is required for reconstruction. As long as the number of colluding malicious aggregators is less than $t$, they cannot reconstruct the functional key or compromise the privacy of the aggregated model.

(iii) Poisoning Attack:
Unlike the 'honest-but-curious' assumption in many schemes, we assume aggregators can be adversarial and may attempt to launch poisoning attacks by returning tampered or random results to disrupt the training process. \textsf{VTSAFL} directly counters this threat through its verifiability feature. Each aggregator must compute a proof of correctness $\Pi_j$ for its partial decryption result $(ct_j^{(k)})$. Clients then run the \texttt{\textsf{VTMCFE.Verify}} algorithm to validate these proofs. If an aggregator provides a malicious or incorrect result, the verification will fail, and its result will be discarded by the clients. This ensures the integrity of the global model aggregation.

(iv) Replay Attack:
An adversary might attempt to replay ciphertexts from previous training rounds to disrupt the current aggregation. To prevent this, our scheme associates a unique label $l^{(k)}$ with each training round $k$. The underlying Multi-Client Functional Encryption (\textsf{MCFE}) primitive ensures that ciphertexts can only be combined for decryption if they possess the same label. Therefore, a ciphertext from a previous round with label $l^{(k-1)}$ will be incompatible with ciphertexts from the current round with label $l^{(k)}$, thwarting replay attacks.

\section{Performance Evaluation}\label{sec:performance}

In this section, we evaluate the proposed \textsf{VTSAFL} framework against the state-of-the-art \textsf{TAPFed} scheme \cite{TAPFed}, focusing on model quality, computational efficiency, and communication overhead. 
Experimental results demonstrate that \textsf{VTSAFL} matches \textsf{TAPFed}'s model accuracy while significantly reducing both running time and communication costs. 
Furthermore, our framework provides the added security benefit of verifiable aggregation.

\begin{table*}[!t]
	\centering
	\caption{Experimental Setting} 
	\label{tab:exp_setup} 
	\ref{tab:exp_setup}
	\begin{tabular}{ll c ccccc}
		\toprule
		Dataset & Samples(train\textbar{}test)/Party & Model Parameters & Training Rounds & Local Epochs & Local Batch \\
		\midrule
		MNIST & 1,000\textbar{}200 & 110,170 & 20 & 10 & 50 \\
		CIFAR10 & 10,000\textbar{}2,000 & 129,242 & 20 & 5 & 50 \\
		\bottomrule
	\end{tabular}
\end{table*}

\begin{table*}[!t]
	\caption{Performance Comparison of Cryptographic Primitives (ms)}
	\label{tab:perf_compare_final_updated}
	\begin{center}
		\footnotesize
		\setlength{\tabcolsep}{3.5pt}
		\begin{tabular}{%
				>{\centering\arraybackslash}p{1.0cm}
				>{\centering\arraybackslash}p{1.3cm}
				>{\centering\arraybackslash}p{1.3cm}
				>{\centering\arraybackslash}p{1.4cm}
				>{\centering\arraybackslash}p{1.3cm}
				>{\centering\arraybackslash}p{1.3cm}
				>{\centering\arraybackslash}p{1.3cm}
				>{\centering\arraybackslash}p{1.4cm}
				>{\centering\arraybackslash}p{1.3cm}
				>{\centering\arraybackslash}p{1.3cm}
			}
			\toprule
			& \multicolumn{4}{c}{\textsf{TAPFed}\cite{TAPFed}} & \multicolumn{5}{c}{Our Scheme} \\
			\cmidrule(lr){2-5} \cmidrule(lr){6-10}
			Dimension ($n$) & \textsf{DKeyGen} & \textsf{Encrypt} (Avg) & \textsf{Partial Decrypt} & \textsf{Combine} & \textsf{DKeyGen} & \textsf{Encrypt} (Avg) & \textsf{Partial Decrypt} & \textsf{Verify} & \textsf{Combine} \\ 
			\midrule
			5   & 0.143 & 0.114 & 0.419 & 0.718 & 0.069 & 0.030 & 0.800 & 0.794 & 0.547\\
			10  & 0.229 & 0.112 & 0.772 & 0.734 & 0.071 & 0.030 & 0.856 & 0.849 & 0.603 \\
			50  & 0.990 & 0.126 & 3.757 & 0.799 & 0.176 & 0.039 & 1.488 & 1.191 & 0.862 \\
			100 & 2.042 & 0.146 & 9.113 & 0.855 & 0.324 & 0.056 & 2.559 & 1.737 & 1.209 \\
			\bottomrule
		\end{tabular}
	\end{center}
\end{table*}

\begin{figure*}[!t]
	\centering
	\subfloat[\textsf{DKeyGen}]{
		\includegraphics[width=0.3\textwidth]{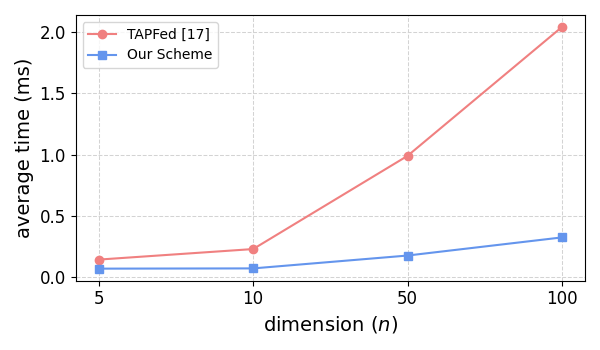}
		\label{fig:keygen}
	}
	\hfill
	\subfloat[\textsf{Encrypt (Avg)}]{
		\includegraphics[width=0.3\textwidth]{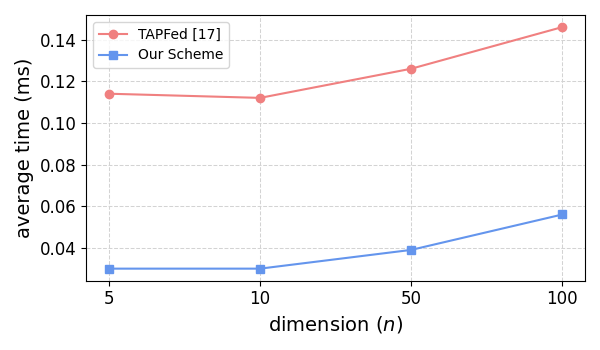}
		\label{fig:encrypt_avg}
	}
	\hfill
	\subfloat[\textsf{Partial Decrypt}]{
		\includegraphics[width=0.3\textwidth]{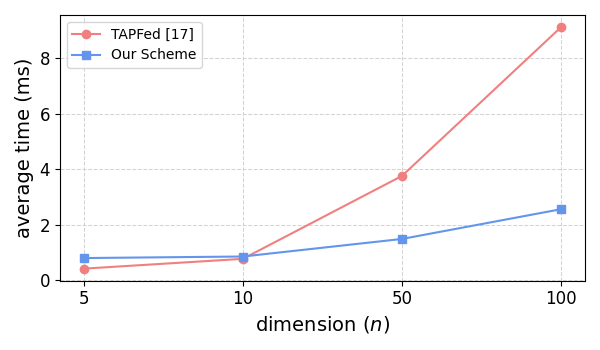}
		\label{fig:partial_decrypt}
	}
	
	\vspace{0.1cm} 
	
	\subfloat[\textsf{Combine Decrypt}]{
		\includegraphics[width=0.3\textwidth]{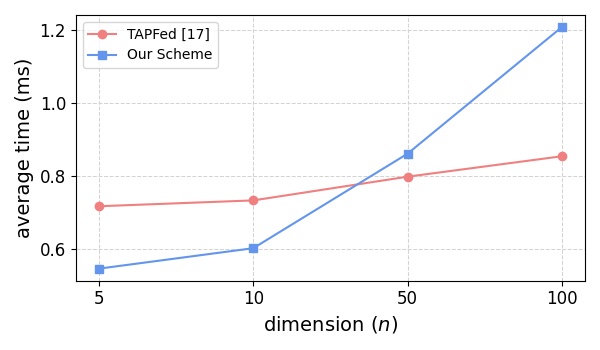}
		\label{fig:combine}
	}
	\hspace{0.03\textwidth}
	\subfloat[\textsf{Verify}]{
		\includegraphics[width=0.3\textwidth]{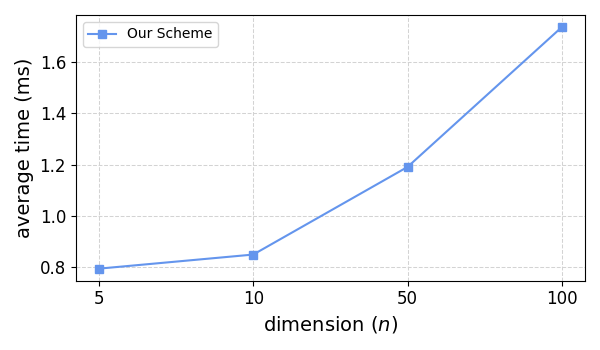}
		\label{fig:verify}
	}
	\hfill
	
	\caption{Performance Comparison of Core Cryptographic Primitives}
	\label{fig:all_performance_charts}
\end{figure*}

\subsection{Experimental Setup}

\textit{Implementation.} We utilized the TensorFlow library for training the federated learning models and implement our framework in python. For a fair comparison, the core cryptographic components of the \textsf{TAPFed} scheme \cite{TAPFed} were re-implemented precisely as described in the original publication. All experiments were performed on a local simulated system. The hardware configuration comprised an Intel(R) Core(TM) i5-13400F CPU, 32GB of DDR4 RAM, and an NVIDIA GeForce RTX 4070s GPU with 12GB of memory.

\textit{Federated Learning Setting.}
Model performance was evaluated using the public MNIST and CIFAR10 datasets. A Keras-based Convolutional Neural Network (CNN) architecture was employed. The model configured for MNIST contained 110,170 parameters, while the CIFAR10 model comprised 129,242 parameters. During each training round, clients first train models on their local datasets, after which the resulting model updates are encrypted and uploaded. A comparative analysis was conducted against the \textsf{TAPFed} scheme \cite{TAPFed}, focusing on model accuracy, training time, and communication overhead. This analysis involved varying both the number of participating clients and the dimensionality of the model vectors.

Our experimental evaluation was structured in two main phases. The initial phase focused on assessing model performance. Here, both our proposed scheme and \textsf{TAPFed} were trained for 20 rounds on the MNIST and CIFAR10 datasets to compare their test accuracies. Relevant hyperparameters and experimental settings are detailed in Table~\ref{tab:exp_setup}. The second phase was designed to evaluate the training time and communication overhead of both schemes as the number of clients varied. Experiments were conducted with the number of clients set to 5, 10, 20, 30, 40, and 50, respectively.

\subsection{Experimental Results}

\begin{figure*}[htbp]
	\centering
	
	\subfloat[]{
		\includegraphics[width=0.23\textwidth]{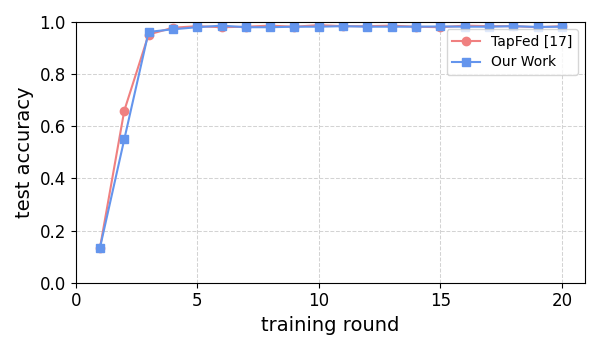}
		\label{fig:Test accuracy}
	}
	\hfill
	\subfloat[]{
		\includegraphics[width=0.23\textwidth]{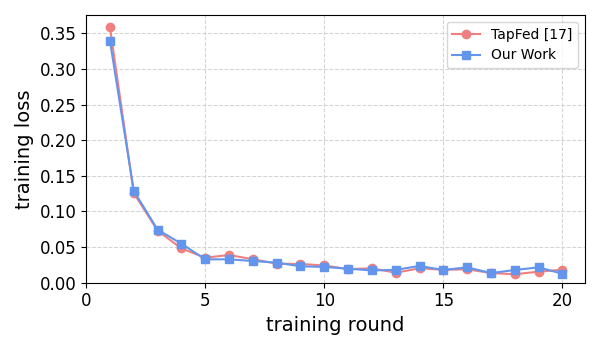}
		\label{fig:Training loss}
	}
	\hfill
	\subfloat[]{
		\includegraphics[width=0.23\textwidth]{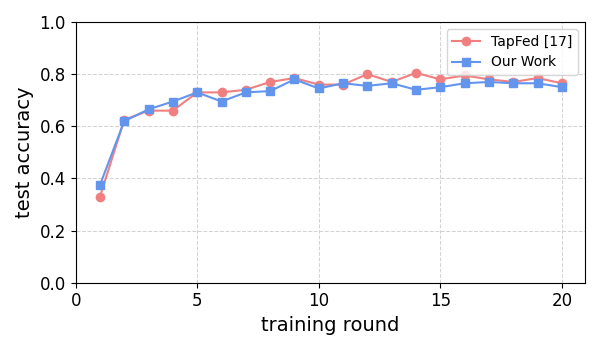}
		\label{fig:Test accuracy CIFAR10}
	}
	\hfill
	\subfloat[]{
		\includegraphics[width=0.23\textwidth]{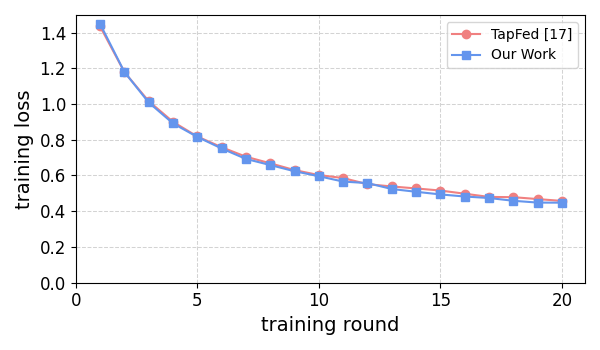}
		\label{fig:Training loss CIFAR10}
	}
	
	\caption{Comparison of test accuracy and training loss on MNIST (a, b) and CIFAR-10 (c, d).}
	\label{fig:combined_results}
\end{figure*}

\begin{figure*}[!t]
	\centering
	
	\subfloat[]{
		\includegraphics[width=0.23\textwidth]{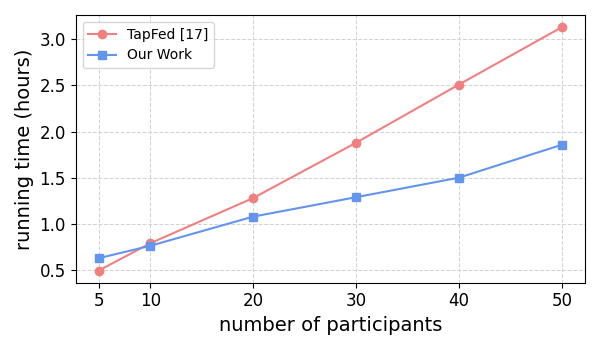}
		\label{fig:mnist_time} 
	}
	\hfill
	\subfloat[]{
		\includegraphics[width=0.23\textwidth]{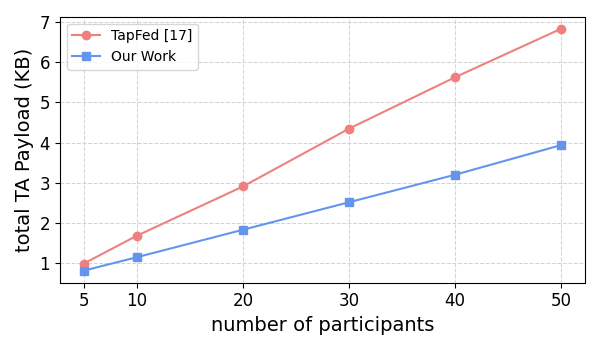}
		\label{fig:mnist_comm}
	}
	\hfill
	\subfloat[]{
		\includegraphics[width=0.23\textwidth]{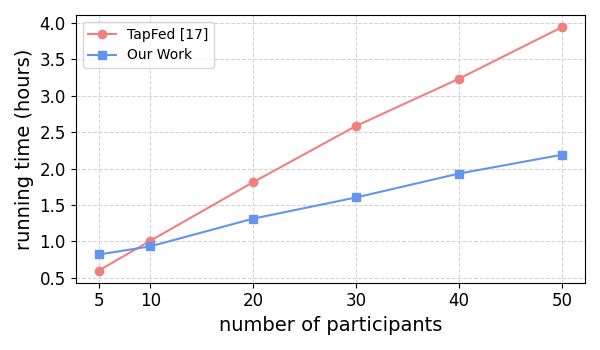}
		\label{fig:cifar_time}
	}
	\hfill
	\subfloat[]{
		\includegraphics[width=0.23\textwidth]{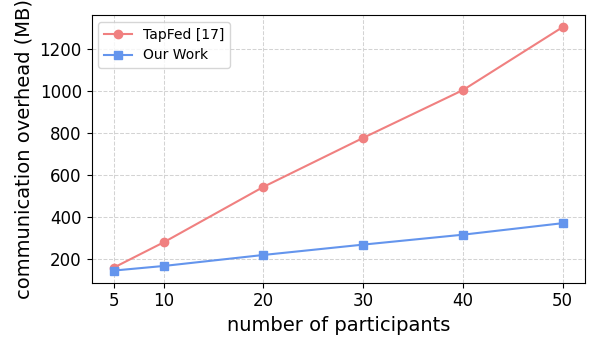}
		\label{fig:cifar_comm}
	}
	
	\caption{Comparison of running time and communication overhead on MNIST (a, b) and CIFAR-10 (c, d).}
	
	\label{fig:running_time_and_comm_all}	
\end{figure*}

\textit{Analysis of Cryptographic Primitives.} To provide a more detailed view, we analyzed the performance of the core cryptographic primitive, as shown in the Table \ref{tab:perf_compare_final_updated} and the Fig.~\ref{fig:all_performance_charts}. Our scheme shows significant efficiency improvement in the \texttt{\textsf{DKeyGen}} and \texttt{\textsf{Partial Decrypt}} stages, which is the main factor to reduce the overall training time. Although our \texttt{\textsf{Combine Decrypt}} is equivalent to \textsf{TAPFed} \cite{TAPFed}, our framework introduces a key \texttt{\textsf{Verify}} step. The time cost of this verification is reasonable, and it provides the necessary guarantee for the correctness of the aggregation results, which is the missing feature of \textsf{TAPFed} \cite{TAPFed}.

\textit{Model Quality and Performance.} We first evaluate the model quality of \textsf{VTSAFL} relative to the baseline. The experimental results show that, compared with standard federated learning, the proposed security aggregation protocol will not adversely affect the key performance indicators, namely learning convergence speed, training loss and test accuracy. As shown in Fig.~\ref{fig:combined_results}, during the whole 20 rounds of federal learning and training, the performance curve of test accuracy and training loss generated by our scheme is very close to that of \textsf{TAPFed}. This result is attributed to the design of \textsf{VTSAFL}, which uses pure password mechanism to protect the exchange of model parameters. By avoiding introducing noise, our method achieves the same model quality as other password based security aggregation baselines.

\textit{Training Time.} We studied the effect of security aggregation method on training time and compared it with \textsf{TAPFed}. As shown in Fig.~\ref{fig:running_time_and_comm_all} (a) and Fig.~\ref{fig:running_time_and_comm_all} (c), the total running time of \textsf{VTSAFL} is significantly lower than that of \textsf{TAPFed}, and the performance gap will expand with the increase of the number of clients. In 50 client scenarios, our framework achieved more than 40\% time reduction on both datasets. Specifically, the running time for MNIST is reduced by 40.6\%, and for the more complex CIFAR10 dataset, the reduction is 44.4\%. This highlights the superior time performance of our scheme in a practical, large-scale FL setting, which directly benefits from the highly efficient \textsf{DKeyGen} and \textsf{Decrypt} primitives analyzed previously.

\begin{table} [ht]
	\caption{Comparison of Communication Cost per Round}
	\label{tab:comm_cost}
	\begin{center}
		\begin{tabular}{%
				>{\centering\arraybackslash}p{1.6cm}
				>{\centering\arraybackslash}p{2cm}
				>{\centering\arraybackslash}p{1.3cm}
				>{\centering\arraybackslash}p{2.2cm}
			}
			\toprule
			& \textsf{DKeyGen} & \textsf{Encrypt} & \textsf{ShareDecrypt} \\
			\midrule
			\textsf{TAPFed}\cite{TAPFed}& $(n+1)|\mathbb{Z}_p|$ & $2|G|$ &$(n+2)|G|$ \\
			\textbf{our work} & $2|G|+|\mathbb{Z}_p|$ & $|G|$ & $5|G|+2|\mathbb{Z}_p|$\\
			\bottomrule
		\end{tabular}
	\end{center}
\end{table}

	\textit{Communication Overhead.} In order to evaluate the communication efficiency of our scheme, we measured the transmission overhead in each round of training through experimental comparison. As indicated in Table~\ref{tab:comm_cost} and illustrated in Fig.~\ref{fig:running_time_and_comm_all} (b) and Fig.~\ref{fig:running_time_and_comm_all} (d), \textsf{VTSAFL} achieves a substantially lower transmission payload. For the client-side upload, our framework reduces the payload by 50.1\% for MNIST and 50.8\% for CIFAR10 in the 50-client scenario. While this client-upload cost scales linearly with $n$ for both schemes, \textsf{VTSAFL}'s scalability is far superior. Furthermore, this advantage is compounded by other protocol phases. As detailed in Table~\ref{tab:comm_cost}, the communication costs for key phases in \textsf{TAPFed} (i.e., \textsf{DKeyGen} and \textsf{ShareDecrypt}) are also proportional to $n$. In sharp contrast, the corresponding communication phases in \textsf{VTSAFL} (e.g., DS-to-Aggregator) have a {constant cost that is independent of $n$.
	
	This performance advantage underscores the superior scalability of our framework. As shown in Fig.~\ref{fig:running_time_and_comm_all}, both the total running time and communication overhead of \textsf{VTSAFL} scale much more efficiently compared to \textsf{TAPFed} \cite{TAPFed}. This confirms that our framework is exceptionally well-suited for large-scale federated learning tasks, offering a practical path to high efficiency and strong security at scale.
	
	\section{Conclusion}\label{sec:conclusion}
	
	In this article, we propose a new privacy preserving federated learning security aggregation framework \textsf{VTSAFL}, to address the security vulnerabilities of existing solutions in the face of malicious aggregators. Based on our partial decryption verifiable threshold multi client function encryption scheme, the core of \textsf{VTSAFL} is constructed. The framework can not only resist the gradient reasoning and data leakage attacks in the existing schemes, but also enable the client to verify the correctness of the aggregation results. This effectively prevents poisoning attacks from malicious aggregators and ensures the integrity of model training.
	
	At the same time, in order to adapt to the scenario of more participating devices in the internet of things environment and avoid the limitation of the computing power of edge devices, we are committed to improving the computing efficiency and communication overhead of existing solutions. Experimental results show that \textsf{VTSAFL} has significant advantages in computational performance and communication efficiency compared with existing schemes, and achieves the same model performance. For internet of things scenarios with a large number of devices, the scheme has good scalability.
	
	For future work, we plan to explore several directions. Our goal is to extend the application of \textsf{VTSAFL} to a broader cross device joint learning environment to verify its practicability and performance under different network and hardware conditions.

	\section*{Acknowledgement}
	
	This work was supported in part by the National Natural Science Foundation of China under Grant 62372103, the
	Natural Science Foundation of Jiangsu Province under Grant BK20231149 and the Jiangsu Provincial Scientific Research Center of Applied Mathematics under Grant BK20233002.

\begin{IEEEbiography}[{\includegraphics[width=1in,height=1.25in,clip,keepaspectratio]{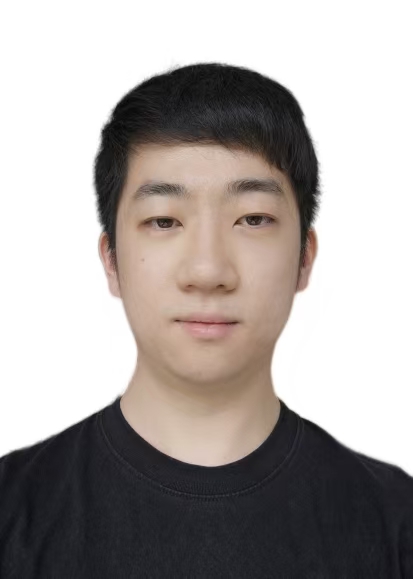}}]{Minjie Wang} received the B.E. degree from School of Computer Science, Nanjing University of Posts and Telecommunications, Nanjing, China, in 2023. He is currently working toward the the master’s degree with the School of Cyber Science and Engineering, Southeast University, Nanjing, China. His main research directions include functional encryption and privacy computing.
\end{IEEEbiography}

\begin{IEEEbiography}[{\includegraphics[width=1in,height=1.25in,clip,keepaspectratio]{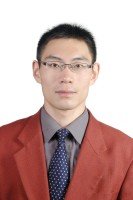}}]{Jinguang Han}(Senior Member, IEEE) received the PhD degree from the University of Wollongong, Australia, in 2013. He is a professor with the School of Cyber Science and Engineering, Southeast University, China. His research focuses on access control, cryptography, cloud computing, and privacy-preserving systems. He served as a Program Committee Co-Chair for ProvSec’2016, FCS’2019, and SPNCE’2020; and a ProgramCommitteeMemberforseveralconferences, including SecureCom’2023, ISC’2022, PST’2021, ESORICS’2020, and ICICS’2019.\end{IEEEbiography}

\begin{IEEEbiography}[{\includegraphics[width=1in,height=1.25in,clip,keepaspectratio]{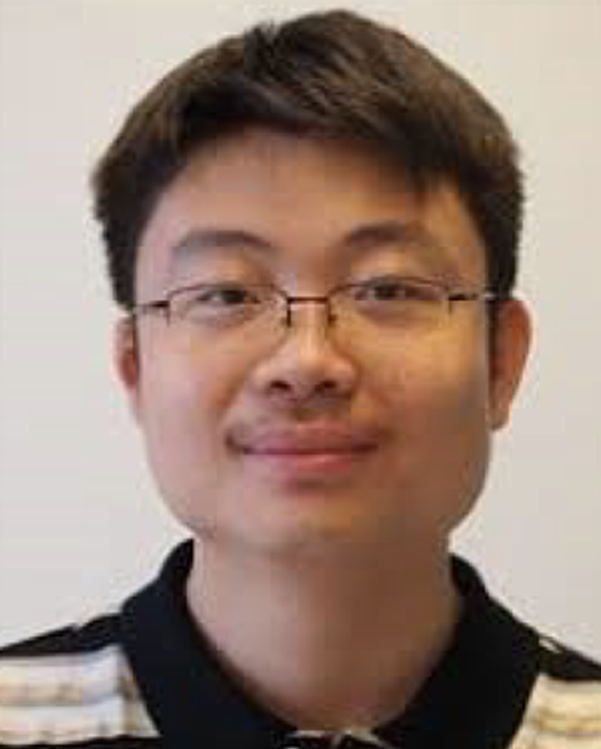}}]{Weizhi Meng}(Senior Member, IEEE) received the Ph.D. degree in computer science from the City University of Hong Kong. He is currently a Full Professor with the School of Computing and Communications, Lancaster University, U.K., and an Adjunct Faculty Member with the Department of Applied Mathematics and Computer Science, Technical University of Denmark, Denmark. His primary research interests include blockchain technology, cyber security, and artificial intelligence in security, including intrusion detection, blockchain applications, smartphone security, biometric authentication, and the IoT security. He was a recipient of Hong Kong Institution of Engineers (HKIE) Outstanding Paper Award for Young Engineers/Researchers in both 2014 and 2017. He also received the IEEE ComSoc Best Young Researcher Award for Europe, Middle East, and Africa Region (EMEA) in 2020, and the IEEE ComSoc Communications and Information Security (CISTC) Early Career Award in 2023.\end{IEEEbiography}

\end{document}